\documentclass[letterpaper,titlepage,12pt]{article}
\usepackage{hyperref}

\usepackage{amssymb,amsmath,amsfonts}
\usepackage{nicefrac}
\usepackage{scalefnt}
\usepackage[titles]{tocloft}
\usepackage{sectsty}
\allsectionsfont{\bf \scalefont{.7} \selectfont}
\subsectionfont{\bf \scalefont{.85} \it \selectfont}
\subsubsectionfont{\bf \scalefont{1} \it \selectfont}
\usepackage{epsfig}
\usepackage[T1]{fontenc}
\usepackage{lmodern}
\usepackage{bibspacing}

\makeatletter
\def\@seccntformat#1{\csname the#1\endcsname.\quad}
\makeatother

\def\blfootnote{\xdef\@thefnmark{}\@footnotetext}

\long\def\symbolfootnote[#1]#2{\begingroup%
\def\thefootnote{\fnsymbol{footnote}}\footnote[#1]{#2}\endgroup}

\makeatletter
\renewcommand{\@dotsep}{4.5} 
\makeatother
\def\clock{{\count0=\time
           \divide\count0 60
           \ifnum\count0<10 0\fi\the\count0
           \multiply\count0 -60 \advance\count0 \time
           :\ifnum\count0<10 0\fi \the\count0
         }}
\newcommand{\timestamp}{{\small\vbox{\hbox{\tt\jobname.tex}
\hbox{\the\day/\the\month/\the\year, \clock}}}}

\newcommand {\slsh} [1] {\not{\hbox{\kern-2pt${#1}$}}}
\newcommand{\drawsquare}[2]{\hbox{%
\rule{#2pt}{#1pt}\hskip-#2pt%  left vertical
\rule{#1pt}{#2pt}\hskip-#1pt%  lower horizontal
\rule[#1pt]{#1pt}{#2pt}}\rule[#1pt]{#2pt}{#2pt}\hskip-#2pt%upper horizontal
\rule{#2pt}{#1pt}}% right vertical

\newcommand{\Yfund}{\raisebox{-.5pt}{\drawsquare{6.5}{0.4}}}%  fund
\newcommand{\Yasymm}{\raisebox
{-3.5pt}{\drawsquare{6.5}{0.4}}\hskip-6.9pt%
                      \raisebox{3pt}{\drawsquare{6.5}{0.4}}%
                     }%  antisymmetric second rank
\newcommand{\Ysymm}{\Yfund\hskip-0.4pt%
                     \Yfund}%  symmetric second rank

\newcommand\T{\rule{0pt}{2.5ex}}

\newcommand\B{\rule[-1.7ex]{0pt}{0pt}}

\def\drawbox#1#2{\hrule height#2pt
         \hbox{\vrule width#2pt height#1pt \kern#1pt
               \vrule width#2pt}
               \hrule height#2pt}

\def\Asym#1#2{\vcenter{\vbox{\drawbox{#1}{#2}
               \kern-#2pt       % line up boxes
               \drawbox{#1}{#2}}}}

\def\SU{\operatorname{SU}}
\def\U{\operatorname{U}}
\def\bdot{\huge{\textbf{.}}}

\def\AA{{\cal A}}
\def\BB{{\cal B}}
\def\CC{{\cal C}}

\def\II{{\cal I}}

\def\KK{{\cal K}}

\def\MM{{\cal M}}
\def\NN{{\cal N}}
\def\OO{{\cal O}}
\def\PP{{\cal P}}

\def\SS{{\cal S}}
\def\TT{{\cal T}}

\def\d{{\partial}}
\def\di{{\rm d}}

\def\chc{{{\rm ch}_c}}
\def\Chc{{{\rm Ch}_c}}
\def\chid{{{\rm ch}_{\mathbb{I}}}}
\def\Chid{{{\rm Ch}_{\mathbb{I}}}}
\def\rrangle{{\rangle\rangle}}

\newcommand{\beq}{\begin{equation}}
\newcommand{\eeq}{\end{equation}}
\newcommand{\ba}{\begin{array}}
\newcommand{\ea}{\end{array}}
\newcommand{\bea}{\begin{eqnarray}}
\newcommand{\eea}{\end{eqnarray}}
\newcommand {\ber}{\begin{eqnarray*}}
\newcommand {\eer}{\end{eqnarray*}}

\newcommand{\Dslash}{\,{\raise.15ex\hbox{/}\mkern-12mu D}}

\newcommand{\Z}{\mathbb{Z}}

\newcommand{\R}{\mathbb{R}}

\newcommand{\tr}{\mathop{{\rm Tr}}}

\numberwithin{equation}{section}
\begin{document}

\begin{titlepage}
\begin{flushright}
CPHT-RR165.1207\\
\end{flushright}
\vskip 2.2cm
\begin{center}
\scalebox{2.2}[2.1]{\footnotesize Non-Supersymmetric Seiberg Duality,}
\vskip 0.4cm
\scalebox{2.2}[2.1]{\footnotesize Orientifold QCD and Non-Critical Strings}
\vskip 1.4cm
{\it Adi Armoni,$^{\dagger}$ Dan Isra\"el,$^{\ddagger}$ 
Gregory Moraitis\,$^{\dagger}$ and Vasilis Niarchos\,$^{\natural}$}
\symbolfootnote[4]{a.armoni@swansea.ac.uk, israel@iap.fr, 
pygm@swansea.ac.uk, niarchos@cpht.polytechnique.fr}\\
\vskip 0.5cm
\medskip
{$^{\dagger}$ Department of Physics, Swansea University \\
Singleton Park, Swansea, SA2 8PP, UK}\\
\smallskip
{$^{\ddagger}$ GRECO, Institut d'Astrophysique de Paris,
98bis Bd Arago, 75014 Paris, France}\\
{Unit\'e mixte de Recherche 7095, CNRS -- Universit\'e Pierre et Marie Curie}\\
\smallskip
{$^{\natural}$ Centre de Physique Th\'eorique, \'Ecole Polytechnique,
91128 Palaiseau, France}\\
{Unit\'e mixte de Recherche 7644, CNRS}\\

\end{center}
\vskip .1in

\baselineskip 20pt
\date{}

\vskip 0.25cm \noindent
We propose an electric-magnetic duality and conjecture an exact
conformal window for a class of non-supersymmetric
$U(N_c)$ gauge theories with fermions in the (anti)symmetric representation
of the gauge group and $N_f$ additional scalar and fermion flavors. 
The duality exchanges $N_c \to N_f -N_c \mp 4$ leaving $N_f$ invariant, 
and has common features with Seiberg duality in $\NN=1$ SQCD 
with $SU$ or $SO/Sp$ gauge group. At large $N$ the duality holds 
due to planar equivalence with $\NN=1$ SQCD.  At finite $N$ we embed 
these gauge theories in a setup with D-branes and orientifolds in a 
non-supersymmetric, but tachyon-free, non-critical type 0B string theory  
and argue in favor of the duality in terms of boundary and crosscap state 
monodromies as in analogous supersymmetric situations.  
One can verify explicitly that the resulting duals have matching global anomalies. 
Finally, we comment on the moduli space of these gauge theories and
discuss other potential non-supersymmetric examples that could exhibit similar
dualities.

\vskip .5cm
\noindent
January, 2008
\end{titlepage} \vfill\eject

\tableofcontents

%%%%%%%%%%%%%%%%%%%%%%%%%%%%%%%%%%%%%%%%%
\section{Introduction}
\label{introduction}

%%%%%%%%%%%%%%%%%%%%%%
\vskip -.2cm
\subsection{Setting the stage}

Seiberg duality \cite{Seiberg:1994pq} is an impressive statement 
about the infrared dynamics of strongly coupled gauge theories. It 
states that two gauge theories, which are distinct in the ultraviolet (UV), 
flow in the infrared (IR) to the same fixed point. In the prototypical example
of $\NN=1$ SQCD with gauge group $SU(N_c)$ and $N_f$ flavors, the
low energy dynamics in the conformal window ($3N_c/2<N_f<3N_c$) is 
described by a non-trivial scale invariant theory of interacting quarks 
and gluons, which has a dual formulation in terms of a ``magnetic'' theory
with gauge group $SU(N_f-N_c)$ and the same of number of flavors. 
Currently there is no proof of Seiberg duality, but in supersymmetric cases
($e.g.$ $\NN=1$ SQCD) there is overwhelming evidence 
both from field theory \cite{Seiberg:1994pq} and string theory 
\cite{Elitzur:1997fh,Evans:1997hk}. 

In field theory, besides 't~Hooft anomaly matching,
which is not based on supersymmetry, one can perform a number of 
non-trivial consistency checks that rely on the power of supersymmetry; 
in particular, holomorphy and the properties of the superconformal algebra
(see \cite{Intriligator:1995au} for a review). 

In string theory, one embeds the gauge theory of interest in a D-brane setup 
in a ten-dimensional type II superstring vacuum 
(with fivebranes in flat space -- for a review see \cite{Giveon:1998sr} 
and references therein -- or in near-singular Calabi-Yau compactifications 
\cite{Ooguri:1997ih}), or lifts to M-theory 
\cite{Witten:1997sc,Hori:1997ab,Witten:1997ep,Brandhuber:1997iy,Hori:1998iw}. 
These situations include extra degrees of freedom, but by going to a convenient region 
of moduli space (alas, typically not the one of direct interest for the 
gauge theory) the description simplifies and one can draw interesting
conclusions. 

More recently, it has been understood how to obtain 
a useful, controllable embedding of interesting SQCD-like theories in the corner
of parameter space most relevant for the gauge theory. For example, in the case
of $\NN=1$ SQCD (with a quartic coupling as we will see in section 
\ref{sec:embedding}), this description involves $N_c$ D3- and $N_f$ 
D5-branes in non-critical type IIB superstrings on 
\beq
\label{introaa}
\R^{3,1}\times SL(2)_1/U(1) 
~,
\eeq
where $SL(2)_1/U(1)$ is a supersymmetric coset of the $SL(2,\R)$ 
Wess-Zumino-Witten (WZW) model at level 1 \cite{Fotopoulos:2005cn}. 
String theory in~\eqref{introaa} arises within the ten-dimensional setup 
that realizes $\NN=1$ SQCD (with two orthogonal NS5-branes), in a 
suitable near-horizon decoupling limit that takes properly into account 
the backreaction of the NS5-branes \cite{Giveon:1999px}. Since the setup 
provided by~\eqref{introaa} admits an exact worldsheet 
conformal field theory (CFT) formulation, an explicit analysis 
of perturbative string theory is possible in this case. 

It has been proposed in \cite{Murthy:2006xt} that Seiberg duality for 
$\NN=1$ SQCD can be understood in the non-critical string context in 
terms of D-brane monodromies. We will review this argument
in detail in section \ref{sec:embedding} with some important 
new elements that modify some of the basic points in the analysis of
\cite{Murthy:2006xt}. We will find this implementation of the duality 
particularly useful in this paper, because in contrast 
with the non-critical string considered here, the critical ten-dimensional 
description suffers from a closed string tachyon instability and its M-theory 
lift is less understood. 

It is an interesting and compelling question whether similar phenomena, $e.g.$ 
the appearance of a conformal window or the existence of Seiberg dual theories,
can arise and be properly understood in non-supersymmetric gauge theories.
Here are some known facts about this question.

The presence of a conformal window in QCD, more specifically $SU(3)$ 
Yang-Mills theory\footnote{Similar statements can be made also for any 
number of colors $N_c>3$.} with $N_f$ Dirac fermions in the fundamental 
representation was argued many years ago in \cite{Banks:1981nn}.
The conclusions of \cite{Banks:1981nn} are based on a perturbative analysis of the
two-loop beta function, which predicts a conformal window for
$N_f^*<N_f<33/2$. Ref.\ \cite{Appelquist:1996dq} estimated the lower
bound of the conformal window in $SU(N_c)$ QCD 
with $N_f$ flavors at $N_f^*\simeq 4N_c$ ($i.e.$ $N_f \simeq 12$
for $N_c=3$). Lattice Monte Carlo studies \cite{Kogut:1987ai,Brown:1992fz} 
of QCD with $N_f$ flavors suggest $N_f^*>8$. In search of a QCD Seiberg 
dual, 't~Hooft anomaly matching was examined without definite conclusion 
in \cite{Terning:1997xy}. 

Seiberg duality in the context of non-supersymmetric models has been
discussed also in theories that are connected to $\NN=1$ SQCD by soft
supersymmetry breaking \cite{Aharony:1995zh,Evans:1995rv,Cheng:1998xg} 
or by a discrete projection that breaks supersymmetry. The first attempt to 
construct a non-supersymmetric Seiberg dual in the large $N$ limit by discrete 
projection was made by Schmaltz in \cite{Schmaltz:1998bg} using the orbifold 
projection of \cite{Kachru:1998ys}. Schmaltz proposed a $U(N_c)\times U(N_c)$ 
``orbifold'' theory as a candidate for the electric theory. Seiberg duality was 
expected to follow in the large $N$ limit as a consequence of a ``planar equivalence'' 
between the parent and daughter theories of the projection. The validity of this 
proposal relies, however, on the presence of an unbroken global $\Z_2$ 
symmetry \cite{Kovtun:2004bz,Unsal:2006pj}, which is not always guaranteed. 
In the string realization of the gauge theory, this requirement translates to a very simple 
condition: the condition that the closed string background is tachyon-free 
\cite{Armoni:2007jt}. Hence, the string realization of the gauge theory in 
\cite{Schmaltz:1998bg} with D-branes and NS5-branes in the {\em tachyonic} type 
0B string theory \cite{Armoni:1999gc} suggests that the necessary $\Z_2$ symmetry is in 
fact broken in this case. It was therefore proposed in \cite{Angelantonj:1999qg} 
that a gauge theory that lives on branes of a {\em non-tachyonic} type 0$'$B string theory
\cite{Sagnotti:1995ga,Sagnotti:1996qj} (the Sagnotti model) would be a better candidate 
for an electric theory. 

This logic leads us naturally to the ``orientifold'' gauge theories of \cite{Armoni:2003gp} 
(for a review see \cite{Armoni:2004uu}). These are QCD-like, non-supersymmetric $U(N_c)$
gauge theories with fermions in the (anti)symmetric representation of the gauge group
and $N_f$ scalar and fermion flavors. In fact, for $N_c=3$ and $N_f=0$ the theory with
fermions in the antisymmetric representation $is$ QCD with one flavor. Moreover,
these theories have been argued, in the large $N$ limit, to be planar equivalent to 
$\NN=1$ SQCD. So we know to leading order in $1/N$ that these theories have the 
same structure as $\NN=1$ SQCD -- in particular, they have a conformal window at 
$\frac{3}{2}N_c<N_f<3N_c$ and exhibit Seiberg duality. Our aim in this paper
is to extend this picture beyond the large $N$ regime, where planar equivalence is lost,
and to make some exact predictions about the conformal window and Seiberg duality
at finite $N$. A specific result is the prediction for a finite $N$ conformal window at
$\frac{3}{2}N_c\mp\frac{20}{3}\leq N_f\leq 3N_c\pm \frac{4}{3}$ with $N_c > 5$. The 
plus/minus signs refer to the specifics of the orientifold projection.
The plan of the paper is as follows.

%%%%%%%%%%%%%%%%%%%%%%%
\subsection{Overview of the paper}

In the first part of section 2 we review known facts about the orientifold field theories 
of interest, namely their definition and symmetries. We consider two models, one 
with the ``gaugino'' in the antisymmetric representation of the $U(N_c)$ gauge group
and another with the ``gaugino'' in the symmetric representation of the gauge group.
We will call the first model OQCD-AS and the second OQCD-S. We will refer to any of
these theories collectively as OQCD. In both cases, there are $N_f$ quark multiplets 
in the fundamental and anti-fundamental of the gauge group. The matter content of 
both theories is summarized in table \ref{tabelectric}.

In the second part of section 2 we present the definition and symmetries of a 
corresponding set of theories, which we will claim are the magnetic duals of 
OQCD-AS and OQCD-S. The matter content of these models is summarized in 
table \ref{tabmagn}.

Our primary motivation for the proposal of this non-supersymmetric electric-magnetic 
duality comes from string theory. In section 3 we explain how we can embed
both the electric and magnetic descriptions of OQCD in a highly curved, but  exact type 
0B non-critical string theory background and how we can motivate Seiberg duality
as a statement about boundary and crosscap state monodromies. The setup involves
$N_c$ D3-branes, $N_f$ D5-branes and an O$'$5-plane that projects out the
tachyonic mode of type 0B string theory. The outcome of the string theory setup is 
a proposal for a duality between the electric description of OQCD-AS (resp.\ OQCD-S) 
with gauge group $U(N_c)$ and $N_f$ flavors and the magnetic description with gauge 
group $U(N_f-N_c+4)$ (resp. $U(N_f-N_c-4)$) and the same number of flavors $N_f$.
A similar analysis has been performed for type IIB string theory in \eqref{introaa} with 
D3- and D5-branes \cite{Fotopoulos:2005cn}, giving Seiberg duality for $\NN=1$ 
SQCD with gauge group $SU(N_c)$ \cite{Murthy:2006xt}. Adding O5-planes in the 
type IIB setup gives Seiberg duality for $SO(N_c)$ or $Sp(N_c/2)$ gauge group
\cite{Ashok:2007sf}. We note, however, that the analysis of the D-brane monodromies 
in the present paper (in both cases of SQCD and OQCD) departs significantly from 
those in the latter two references.  

We want to emphasize that non-critical string theory on \eqref{introaa} with an 
O$'$5-plane is forced upon us in an almost unique way. From the analysis of the 
gauge invariant operators of OQCD in the large $N$ limit \cite{Armoni:2004uu} 
we expect a purely bosonic spectrum. This implies, using general ideas from 
holography \cite{'tHooft:1973jz}, that in order to engineer OQCD with a D-brane 
setup in string theory, we have to embed D-branes in a background with a closed 
string spectrum that is also purely bosonic in space-time. At the same time, the 
validity of planar equivalence with $\NN=1$ SQCD at large $N$ requires 
\cite{Kovtun:2004bz,Unsal:2006pj} that a certain discrete symmetry (charge 
conjugation in the present case) is not spontaneously broken. This condition 
translates to a tachyon-free closed string background \cite{Armoni:2007jt}. 
So gauge theory considerations alone imply that we are looking for a string 
theory that has a purely bosonic closed string spectrum without closed string 
tachyons. Besides two dimensional examples (which are clearly irrelevant for
our purposes) the only other examples that are known with these features are
the non-critical string theories of \cite{Israel:2007nj}, a close cousin of which is 
the theory we are discussing in section 3.\footnote{The Sagnotti model in ten 
dimensions is also a theory with a tachyon-free, purely bosonic closed string 
spectrum, but includes an additional open string sector with spacefilling D9-branes.}

In section 4 we collect the evidence for the proposed duality and discuss
its limitations. The evidence in favor of our proposal includes:
\begin{itemize}
\item[(1)] Seiberg duality is guaranteed to hold at infinite $N$ because of planar 
equivalence with $\NN=1$ SQCD. This alone fixes the basic features of the 
magnetic duals, $e.g.$ the matter content and the form of the dual Lagrangian.
\item[(2)] The string theory embedding motivates a definite proposal for the
duality at finite $N$. The details of the string theory construction and the 
interpolation between the electric and magnetic descriptions are a hybrid of the 
corresponding setups in the context of $\NN=1$ SQCD with $U(N_c)$ or 
$SO(N_c)$/$Sp(N_c/2)$ gauge groups. Hence, certain arguments that can be 
used in favor of the validity of the duality in those cases suggest (however, do
not rigorously prove) the validity of the duality in OQCD as well.
\item[(3)] Global anomalies can be computed explicitly and 't Hooft anomaly 
matching is found to hold for the proposed duals at any $N$.
\end{itemize}

In section 5 we discuss the implications of our proposal for the IR 
dynamics of OQCD. One of them is a prediction for the precise range
of the conformal window and an emerging picture of the phase structure of the
theory as a function of the number of flavors $N_f$. Another interesting 
question concerns the quantum moduli space of OQCD. The parent 
$\NN=1$ SQCD has a whole space of vacua for $N_f\geq N_c$ parametrized 
by the vacuum expectation values (vevs) of meson and baryon fields.
At finite $N$ quantum corrections lift the classical moduli space of OQCD 
and one is left with a unique vacuum. To demonstrate this we compute 
the one-loop Coleman-Weinberg potential for the vev of the squark fields.

We conclude in section 6 with a brief discussion on possible extensions of this work.

%%%%%%%%%%%%%%%%%%%%%%%%%%%%%%%%%%%%%%%%%%
\section{Duality in non-supersymmetric gauge theory: a proposal}
\label{sec:gaugetheory}

In this section, we propose an IR duality between an ``electric'' 
and a ``magnetic'' version of the non-supersymmetric OQCD gauge theories.
We present the precise definition of these theories and summarize their
most salient features.

%%%%%%%%%%%%%%%%%%%%%%
\vskip -.5cm
\subsection{The electric theory}

\begin{table}[!t]%ht
\begin{center}
\scalebox{1.04}[1.04]{
\begin{tabular}{|c|c|ccc||c|ccc|}
\hline
		\multicolumn{5}{|c||} {$\operatorname {OQCD-{AS}}$} & 
		\multicolumn{4}{c|}{$\operatorname {OQCD-S}$}  \\
\hline \hline
               & $\operatorname U(N_c)$ & $\operatorname{SU}(N_f)$ & $\operatorname{SU}(N_f)$ & 
               $\operatorname U(1)_R$
               & $\operatorname U(N_c)$ & $\operatorname{SU}(N_f)$ & $\operatorname{SU}(N_f)$ & 
               $\operatorname U(1)_R$ \\
\hline
$A_\mu$            & adjoint & \bdot & \bdot & 0 & adjoint & \bdot & \bdot & 0 
		\\
                   & $N_c^2$ & & & &$N_c^2$ & & &
                   \\
\hline
$\lambda$ & $\T\Yasymm\B$    &  \bdot  &  \bdot  &  1 & $\T\Ysymm\B$    &  \bdot  &  \bdot  &  1   
		\\
                   & $\tfrac{N_c(N_c-1)}{2}\B$  & & & &$\tfrac{N_c(N_c+1)}{2}\B$  & & & 
                   \\
                   \hline
$\tilde\lambda$ 
	& $\T\overline\Yasymm\B$  &  \bdot  &  \bdot  &  1 & $\T\overline\Ysymm\B$  &  \bdot  &  \bdot  &  1  
		\\
                   & $\tfrac{N_c(N_c-1)}{2}\B$  & & & &$\tfrac{N_c(N_c+1)}{2}\B$  & & & 
                   \\
\hline\hline
$\Phi$   & $\overline\Yfund$ & $\Yfund$ & \bdot & $\T\frac{N_f-N_c+2}{N_f}$ 
	     & $\overline\Yfund$ & $\Yfund$ & \bdot & $\T\frac{N_f-N_c-2}{N_f}$ 
		\\
                   &   $\bar{N_c}$  &   $N_f$ &   & & $\bar{N_c}$  &   $N_f$ &   &  
                   \\
\hline
$\Psi$ & $\Yfund$ & $\Yfund$ & \bdot & $\T\frac{-N_c+2}{N_f}$ 
		  & $\Yfund$ & $\Yfund$ & \bdot & $\T\frac{-N_c-2}{N_f}$ 
		\\
                   &   $N_c$  &   $N_f$ &  & & $N_c$  &   $N_f$ &   & 
                   \\                   
\hline\hline
$\tilde \Phi$	& $\Yfund$ & \bdot & $\overline\Yfund$  & $\T\frac{N_f-N_c+2}{N_f}$ 
		& $\Yfund$ & \bdot & $\overline\Yfund$  & $\T\frac{N_f-N_c-2}{N_f}$ 
		\\
                   &   $N_c$  &  & $\bar{N_f}$ & &$N_c$  &  & $\bar{N_f}$ &  
                   \\
\hline
$\tilde\Psi$	& $\overline \Yfund$ & \bdot & $\overline \Yfund$  & $\T\frac{-N_c+2}{N_f}$ 
		& $\overline \Yfund$ & \bdot & $\overline \Yfund$  & $\T\frac{-N_c-2}{N_f}$ 
		\\
                   &   $\bar{N_c}$  & & $\bar{N_f}$ & & $\bar{N_c}$  & & $\bar{N_f}$ &  
                   \\
\hline
\end{tabular}}
\vskip .4cm
\caption{\it The matter content of the electric description of OQCD. The left (right) four columns
depict the matter content, representations and $U(1)_R$ quantum numbers of the OQCD theory
in the antisymmetric (symmetric) projection.}
\label{tabelectric}
\end{center}
\end{table}

The electric version of the gauge theory we will discuss here comes into two 
variants: OQCD-AS and OQCD-S. The matter content of these theories is given 
in table \ref{tabelectric}. In both cases, the boson representations are identical 
to the boson representations of the original $U(N_c)$ super-QCD. The difference 
occurs in the fermionic sector. The original gaugino is replaced by a ``gaugino'' in 
either the antisymmetric representation (in OQCD-AS) or in the symmetric 
representation (in OQCD-S). Similarly, the quarks (the ``superpartners'' of the 
``squarks''), although still in bifundamental representations, do not transform 
exactly as the quarks of the original supersymmetric theory. Their representations 
are fixed by their coupling to the gaugino and the quark, namely by the terms
\begin{subequations}
\bea 
\label{gautheaa}
& & {\rm OQCD-AS:} \,\,\,  \lambda _{ [ij] } \Phi ^i _{\alpha} \bar \Psi ^{j \alpha}
+ \tilde \lambda ^{ [ij] } \tilde \Phi _ i ^{\alpha} {\bar {\tilde \Psi}} _{j \alpha} \\
 & & {\rm OQCD-S:} \,\,\, \lambda _{\{ij\}} \Phi ^i _{\alpha} \bar \Psi ^{j \alpha} 
 + \tilde \lambda  ^ {\{ij\}} \tilde \Phi _i ^ {\alpha} {\bar {\tilde \Psi}} _{j \alpha} \, ,
\eea  
\end{subequations}
where $i,j$ are color indices and $\alpha$ is a flavor index. The tree level 
Lagrangian of the theory is inherited from the supersymmetric theory, hence 
it contains the same fields and the same interaction terms as in the supersymmetric 
$U(N_c)$ gauge theory.  Altogether, the Lagrangian looks like a hybrid of bosons 
from the $U(N_c)$ and fermions from the $SO(N_c)$ ($Sp(N_c/2)$) theories. 

Both theories exhibit an anomaly free $SU(N_f) \times SU(N_f) \times U(1)_R$ global 
symmetry. We call the anomaly free axial symmetry $U(1)_R$, although it is not an 
R-symmetry. Moreover, it will be important for our considerations below that 
the baryon $U(1)_B$ symmetry is gauged and that the gauge group is $U(N_c)$ 
instead of $SU(N_c)$. Consequently, there are no baryon operators in the theories 
we will consider.

At the classical level, the model admits a moduli space, parametrized by the vevs 
of scalars, exactly as in the supersymmetric $U(N_c)$ theory. At the quantum level,
and finite $N$, this moduli space is lifted as no supersymmetry is present.
This will be discussed further in section 5. Nevertheless, due to planar equivalence 
\cite{Armoni:2004uu} in the large $N$ limit, both OQCD-AS and OQCD-S become 
equivalent to the $U(N_c)$ electric SQCD theory in the common sector of C-parity 
even states \cite{Unsal:2006pj}. Hence, in this limit the non-supersymmetric effects 
are suppressed and OQCD exhibits the same moduli space as the SQCD theory.

%%%%%%%%%%%%%%%%%%%%%%%%
\subsection{The magnetic theory}

\begin{table}[!t]%!tbp
\begin{center}
\scalebox{.98}[1]{
\begin{tabular}{|c|c|ccc||c|ccc|}
\hline
		\multicolumn{5}{|c||} {$\operatorname {OQCD-{AS}}\enspace(\tilde{N_c}=N_f-N_c+4)$} & 
		\multicolumn{4}{c|}{$\operatorname {OQCD-S}\enspace(\hat{N_c}=N_f-N_c-4)$}  \\
\hline\hline
               & $\T\operatorname U(\tilde{N_c})$ & $\operatorname{SU}(N_f)$ & $\operatorname{SU}(N_f)$ & $\operatorname U(1)_R$ & $\T\operatorname U(\hat{N_c})$ & $\operatorname{SU}(N_f)$ & $\operatorname{SU}(N_f)$ & $\operatorname U(1)_R$ \\
\hline \hline
$A_\mu$            & adjoint & \bdot & \bdot & 0 & adjoint & \bdot & \bdot & 0 \\
                   & $\tilde{N_c^2}$ & & & & $\hat{N_c^2}$ & & & \\
\hline
$\lambda         $ & $\T\Yasymm\B$           &  \bdot  &  \bdot  &  1  & $\T\Ysymm\B$       &  \bdot  &  \bdot  &  1  \\
                   & $\tfrac{\tilde N_c(\tilde N_c-1)}{2}\B$  & & & & $\tfrac{\hat N_c(\hat N_c+1)}{2}\B$  & & & \\
\hline
$\tilde\lambda         $ & $\T\overline\Yasymm\B$  &  \bdot  &  \bdot  &  1  & $\T\overline\Ysymm\B$  &  \bdot  &  \bdot  &  1  \\
                   & $\tfrac{\tilde N_c(\tilde N_c-1)}{2} \B$  & & & & $\tfrac{\hat N_c(\hat N_c+1)}{2}\B$  & & & \\
\hline\hline
$\phi$						 & $\Yfund$ & $\overline\Yfund$ & \bdot & $\T\frac{N_c-2}{N_f}$ & $\Yfund$ & $\overline\Yfund$ & \bdot & $\T\frac{N_c+2}{N_f}$ \\
                   &   $\tilde{N_c}$  &   $\bar{N_f}$ &   & &   $\hat{N_c}$  &   $\bar{N_f}$ &   & \\
\hline
$\psi$						 & $\overline \Yfund$ & $\overline \Yfund$ & \bdot & $\T\frac{N_c-N_f-2}{N_f}$ & $\overline \Yfund$ & $\overline \Yfund$ & \bdot & $\T\frac{N_c-N_f+2}{N_f}$ \\
                   &   $\bar{\tilde{N_c}}$  &   $\bar{N_f}$ &   & &   $\bar{\hat{N_c}}$  &   $\bar{N_f}$ &   & \\
\hline\hline
$\tilde \phi$			 & $\overline\Yfund$ & \bdot & $\Yfund$  & $\T\frac{N_c-2}{N_f}$ & $\overline\Yfund$ & \bdot & $\Yfund$  & $\T\frac{N_c+2}{N_f}$ \\
                   &   $\bar{\tilde{N_c}}$  &  & $N_f$ &  &   $\bar{\hat{N_c}}$  &  & $N_f$ &  \\
\hline
$\tilde\psi$		         &   $\Yfund$ & \bdot & $\Yfund$  & $\T\frac{N_c-N_f-2}{N_f}$ &   $\Yfund$ & \bdot & $\Yfund$  & $\T\frac{N_c-N_f+2}{N_f}$ \\
                   &   $\tilde{N_c}$  &  & $N_f$ &  &   $\hat{N_c}$  &  & $N_f$ &  \\
\hline\hline
M                  &   \bdot & $\Yfund$ & $\overline \Yfund$ & $\T\frac{2N_f-2N_c+4}{N_f}$ &   \bdot & $\Yfund$ & $\overline \Yfund$ & $\T\frac{2N_f-2N_c-4}{N_f}$\\
                   &           & $N_f$ & $\bar{N_f}$ & &      & $N_f$ & $\bar{N_f}$ & \\
\hline
$\chi$           &   \bdot & $\Ysymm$ & \bdot & $\T\frac{N_f-2N_c+4}{N_f}$ &   \bdot & $\Yasymm$ & \bdot & $\T\frac{N_f-2N_c-4}{N_f}$\\
                   &   & $\tfrac{N_f(N_f+1)}{2}$ & & &   & $\tfrac{N_f(N_f-1)}{2}$ & &\\
\hline 
$\tilde\chi$           &   \bdot & \bdot & $\overline \Ysymm$ & $\T\frac{N_f-2N_c+4}{N_f}$ &   \bdot & \bdot & $\overline \Yasymm$ & $\T\frac{N_f-2N_c-4}{N_f}$\\
                   &   & & $\T\tfrac{N_f(N_f+1)}{2}$ & &   & & $\T\tfrac{N_f(N_f-1)}{2}$ &\\
\hline
\end{tabular}}
\vskip .4cm
\caption{\it The matter content of the proposed magnetic description of OQCD-AS (left) and OQCD-S (right), with number of colours $\tilde{N_c}=N_f-N_c+4$ and $\hat{N_c}=N_f-N_c-4$ respectively.}
\label{tabmagn}
\end{center}
\end{table}

The planar equivalence at infinite $N$ raises the possibility of an electric-magnetic 
duality in OQCD even at finite $N$. Our purpose here is to make a definite proposal 
for this finite $N$ duality. We will propose that the Seiberg dual of the $U(N_c)$ 
electric OQCD-AS theory is a $U(N_f-N_c+4)$ magnetic OQCD-AS theory. 
Similarly, the dual of the electric OQCD-S is a $U(N_f-N_c-4)$ magnetic 
OQCD-S theory. At infinite $N$, both magnetic duals become planar equivalent 
to the magnetic $U(N_f-N_c)$ SQCD theory.

The matter content of the proposed magnetic theories is summarized in table 
\ref{tabmagn}. Besides the gauge bosons and the ``gluinos'', this table contains,
as in SQCD, additional fundamental degrees of freedom, which comprise of a 
complex scalar field $M$ (the magnetic meson) and Weyl fermions 
$\chi$, $\tilde \chi$ (the magnetic ``mesinos'').  

The representations of the fermions are fixed by the same rules as in 
the electric case. The representation of the gaugino is antisymmetric in 
OQCD-AS and symmetric in OQCD-S. The mesino representation is 
symmetric in OQCD-AS and antisymmetric in OQCD-S, otherwise the 
anomalies would not match. This is similar to the $SO(N)$ ($Sp(N/2)$) SQCD 
case, where the mesino representation is symmetric when the gaugino is 
antisymmetric (and antisymmetric when the gaugino is symmetric). 

The tree level Lagrangians of the magnetic theories are again inherited, due to
planar equivalence at infinite $N$, from the supersymmetric theory. An important
feature of the magnetic Lagrangians are the meson, ``mesino'' couplings
\begin{subequations}
\bea 
\label{gautheba}
& & {\rm OQCD-AS:} \,\,\,  M^\alpha _\beta \psi_{\alpha k} {\tilde \psi} ^{\beta k}
+ \chi_{\{\alpha \beta \}} \phi^\alpha_k \psi^{\beta k}
+ \tilde \chi^{ \{\alpha \beta \}} \tilde \phi_ \alpha ^{k} \tilde \psi _{\beta k} \\
 & & {\rm OQCD-S:} \,\,\, ~~  M^\alpha_\beta \psi_{\alpha k} {\tilde \psi} ^{\beta k}
+\chi_{[\alpha \beta]} \phi^\alpha_{k} \psi^{\beta k}
+  \tilde \chi^{ [\alpha \beta]} \tilde \phi_ \alpha ^k \tilde \psi _{\beta k}
~.
\eea  
\end{subequations}

%%%%%%%%%%%%%%%%%%%%%%%%%%%%%%%%%%%%%%%%%%%%%%
\section{Embedding in string theory}
\label{sec:embedding}

OQCD is a four-dimensional gauge theory with a purely bosonic spectrum of
gauge invariant operators in the large $N$ limit~\cite{Armoni:2004uu}. This is an 
indication, using general ideas from holography~\cite{'tHooft:1973jz}, 
that in order to engineer OQCD with a D-brane setup in string theory, we should
embed D-branes in a non-supersymmetric vacuum with a purely bosonic closed
string spectrum. At the same time, the validity of planar equivalence with $\NN=1$
SQCD at large $N$ requires~\cite{Kovtun:2004bz,Unsal:2006pj,Armoni:2007jt} the 
absence of closed string tachyons. Higher dimensional backgrounds with these
features are not common in string theory. The only such examples that we are aware of 
are the non-critical examples of~\cite{Israel:2007nj}, which are non-critical variants
of the type 0$'$B string theory in ten dimensions \cite{Sagnotti:1995ga,Sagnotti:1996qj}.
Fortunately, a close cousin of the closed string theories presented in \cite{Israel:2007nj} 
provides the right setup for the string theory embedding of OQCD. Electric-magnetic
duality in this context will be discussed using the techniques of~\cite{Murthy:2006xt},
which were applied in an analogous type IIB construction to $\NN=1$ SQCD. Note, 
however, that we correct some statements made in \cite{Murthy:2006xt}, thus 
obtaining a partially modified picture of electric-magnetic duality in the context of 
non-critical string theory.

We would like to mention in passing that there is a close relation between 
the non-critical embedding of gauge theories that we consider here and the 
more familiar Hanany-Witten constructions~\cite{Hanany:1996ie} in critical 
string theory, with configurations of NS5-branes, D-branes and orientifolds 
in flat space-time (see \cite{Giveon:1998sr} for an extensive review). For $\NN=2$ 
gauge theories this relation has been discussed in detail in \cite{Israel:2005fn}. 
For $\NN=1$ SQCD it has been discussed in \cite{Fotopoulos:2005cn,Murthy:2006xt} 
(see also \cite{Israel:2005zp} for related issues). However, in this paper, we will 
not make explicit use of HW constructions (or their M-theory lifts for that matter), 
since they are embedded in tachyonic type 0 string theory and the required 
orientifold planes do not fully project out the closed string tachyon (more 
details are given at the end of subsection~\ref{closedsector}). On the contrary, 
closed string tachyons are not an issue in the non-critical description. 
Moreover, as always, the non-critical description isolates and captures in 
HW constructions the region of moduli space that is most relevant for the 
gauge theory of interest (for a more detailed discussion of these aspects we 
refer the reader to \cite{Israel:2005fn,Fotopoulos:2005cn,Murthy:2006xt} and 
references therein).

In this section we will focus mostly on the physics and the implications of the 
string theory setup at hand. Many explicit technical details are left for the interested 
reader in appendices~\ref{appreps},~\ref{apploop}.

%%%%%%%%%%%%%%%%%%%%%%%%%%%%%%%%%%
\subsection{OQCD in type 0$'$B non-critical strings}
\label{closedsector}

Our starting point is type 0B string theory on the exact perturbative string background
\begin{equation}
\mathbb{R}^{3,1} \times \frac{SL(2)_1}{U(1)} \, .
\label{noncritback}
\end{equation}
The flat $\R^{3,1}$ factor requires on the worldsheet the usual four free bosons
and their fermionic superpartners. The second factor is captured by an axial 
coset of the $SL(2,\R)$ supersymmetric WZW model at level $k=1$. In target 
space this superconformal field theory describes the Euclidean two-dimensional 
black hole \cite{Elitzur:1991cb,Mandal:1991tz,Witten:1991yr} (commonly known 
as the ``cigar'' geometry). At $k=1$ this geometry has curvatures at the string scale 
and is highly curved, hence the geometric intuition is not always a useful or accurate
guide for the description of this setup. Here are some useful facts about the 
supersymmetric $SL(2)/U(1)$ coset.

It is a superconformal field theory with $\NN=(2,2)$ supersymmetry that has a mirror 
dual formulation as $\mathcal{N}=2$ Liouville theory~\cite{fzz,Kazakov:2000pm,Hori:2001ax}. 
An important aspect of this duality is the presence of a non-perturbative winding 
condensate in the CFT. Far from the tip of the cigar, the coset CFT is well approximated 
by the free theory $\mathbb{R}_{\sqrt{2}} \times U(1)_1$, $i.e.$ a linear dilaton 
field $\rho$ with background charge $Q=\sqrt{2}$, a free boson $X$ at the self-dual 
radius\footnote{Hereafter we set $\alpha'=2$.} $R=\sqrt{2}$ and their supersymmetric 
partners, a Dirac fermion with left- and right-moving components 
$\psi^\pm=\psi^\rho\pm i \psi^\textsc{x}$ and 
$\tilde{\psi}^\pm=\tilde{\psi}^\rho\pm i \tilde{\psi}^\textsc{x}$. In terms of this free 
field description, the ${\mathcal N} =2$ Liouville potential reads:
\begin{equation}
\delta \mathcal S_\textsc{l}  (\mu,\bar \mu) = \frac{i}{2\pi} \int \di^2 z \left[ \mu \,
\psi^+ \tilde{\psi}^+ e^{-\frac{\rho+i(X_\textsc{l}-X_\textsc{r})}{\sqrt{2}}} 
+ \bar{\mu}\, \psi^- \tilde{\psi}^- e^{-\frac{\rho-i(X_\textsc{l}-X_\textsc{r})}{\sqrt{2}}}  \right]
\, .
\label{liouvpot}
\end{equation}
$X_{\rm L}$, $X_{\rm R}$ denote the left- and right-moving parts of $X(z,\bar z)$.

In the case at hand $k=1$, and the Liouville coupling constant $\mu$ is given by 
mirror symmetry in terms of the effective string coupling constant 
as follows:\footnote{The relation found in~\cite{Giveon:2001up} needs to be 
renormalized in the limit $k\to 1^+$. See also \cite{Ashok:2005py}. For convenience,
we will denote $\mu^{{\rm ren}; \, k=1}$ simply as $\mu$ in what follows.} 
\begin{equation}
\mu^{{\rm ren} ; \, k=1} : = \lim_{\epsilon\to 0^+} \epsilon^{-\tfrac{1}{2}} \mu^{k=1+\epsilon} 
=  \frac{2}{g_{\rm eff}}
\label{renorm}
\end{equation}
According to this relation, when $\mu \to 0$, the background becomes strongly coupled.
A more detailed description of the supersymmetric $SL(2,\R)/U(1)$ coset and 
$\NN=2$ Liouville theory can be found for example in \cite{Israel:2007si} 
and references therein.

The type 0B theory has a closed string tachyon. In order to project it out of 
the physical spectrum we can use, like in ten dimensions 
\cite{Sagnotti:1995ga,Sagnotti:1996qj}, a spacefilling orientifold; in more 
technical terms, the orientifold should be spacefilling in $\mathbb{R}^{3,1}$ 
and B-type in the $SL(2)/U(1)$ super-coset \cite{Israel:2007nj}. We now proceed 
to discuss the physics of this orientifold and its implications in more detail.

\subsubsection*{The closed string sector}

From the worldsheet point of view, type 0B string theory on~(\ref{noncritback}) 
is described by the tensor product of a free $\mathbb{R}^{3,1}$ super-CFT, the 
usual ghosts and super-ghosts and the super-coset $SL(2)/U(1)$, with 
left-right symmetric boundary conditions for the worldsheet fermions. 
Physical states have delta-function normalizable wavefunctions in the 
coset ($i.e.$ belong to the continuous representations, see appendix \ref{appreps}); 
their left- and right-moving conformal weights, for radial momentum $P$, read
\begin{subequations}
\begin{align}
\Delta & = \frac{1}{2} p_\mu p^\mu + P^2 + \frac{(n+w)^2}{4}+N -\frac{1}{4} = 0~,\\
\bar \Delta & = \frac{1}{2} p_\mu p^\mu + P^2 + \frac{(n-w)^2}{4}+\bar N -\frac{1}{4} = 0~,
\end{align}
\end{subequations}
where $n$ (resp.\ $w$) is the momentum (resp.\ the winding) around the compact direction 
of the coset.\footnote{Only the former is preserved by the interactions taking place near 
the tip of the cigar.} $N$, $\bar N$ are the left-, right-moving oscillator levels and $p_\mu$
is the flat space four-dimensional momentum.
%Unlike type II strings in the same background (see $e.g.$~\cite{Murthy:2003es})
%the GSO projection does not act on the momentum and winding. 
From the torus partition function (see appendix~\ref{apploop}) one finds the 
following spectrum of lowest lying modes (see also~\cite{Murthy:2003es}):
\begin{itemize}
\item A real tachyon with $m^2 = -1/2$ for $n=w=0$ in the NS$_-$NS$_-$ sector.  
\item A pair of complex massless scalars, for $(n=\pm 1,w=0)$ and $(n=0,w=\pm 1)$, 
in the NS$_-$NS$_-$ sector.
\item A real massless RR scalar and a massless self-dual RR two-form in six 
dimensions from the R$_+$R$_+$ sector, plus another real scalar and an 
anti-self-dual two-form from the R$_-$R$_-$ sector. From the four-dimensional 
perspective one gets two complex scalars and two vectors.  
\end{itemize}
The lowest modes in the NS$_+$NS$_+$ fields $(g_{\mu \nu},B_{\mu \nu},\Phi)$ 
are massive, with mass squared $m^2 = 1/2$. The positive
mass shift is due to the presence of a linear dilaton.

In order to obtain a tachyon-free spectrum, we need to perform a specific 
orientifold projection (we refer the reader to~\cite{Israel:2007si} for a 
general analysis of orientifolds in $SL(2)/U(1)$). The orientifold that turns out
to be appropriate for our present purposes is very similar to the one utilized 
in~\cite{Israel:2007nj}, and is defined by the following action 
on primary states:\footnote{This orientifold is B-type in $SL(2)/U(1)$, 
whereas the orientifold appearing in~\cite{Israel:2007nj} was A-type. 
A-type boundary conditions are Dirichlet boundary conditions for the $U(1)$ 
boson $X$ whereas B-type are Neumann.}
\begin{equation}
\mathcal P = \Omega (-)^{Q_R}\, : ~ |n,w\rangle \otimes |0\rangle_{NS} 
\longrightarrow (-)^{n+w+1} |n,-w\rangle \otimes |0\rangle_{NS}\, ,
\label{extpar}
\end{equation}
where $\Omega$ is the standard worldsheet parity. In this definition, $Q_R$
is the charge of the left-moving $U(1)_R$ worldsheet symmetry 
\begin{equation}
Q_R  =\oint \frac{{\rm d}z}{2\pi i} \left(i\psi^\rho \psi^\textsc{x}+ i\sqrt{2}\partial X\right) 
=  F + n+w  \, ,  
\label{wsRcharge}
\end{equation}
where $F$ is the left-moving worldsheet fermion number. 
As with the critical type 0$'$B theory defined 
in~\cite{Sagnotti:1995ga,Sagnotti:1996qj}, the worldsheet parity \eqref{extpar} 
acts on the GSO invariant states of the corresponding type IIB model simply as 
$\Omega$. 

The closed string tachyon is odd under $\PP$, hence it is projected out. 
The resulting theory has a tachyon-free spectrum with a Hagedorn density  
of states, but no spacetime fermions. The invariant massless physical states are
\begin{itemize}
\item A complex massless scalar from the states $|\pm 1,0\rangle \otimes |0\rangle_{NS}$ 
in the NS$_-$NS$_-$ sector.  
\item A real massless scalar from $(|0,+1\rangle + |0,-1\rangle) \otimes |0\rangle_{NS}$ 
in the NS$_-$NS$_-$ sector.  
\item A four-dimensional massless vector and a pair of massless real 
scalars from the R$_+$R$_+$ sector. 
\end{itemize}
Compared to the type IIB case, instead of having two complex massless scalars in the 
NSNS sector one has one complex and one real scalar. The missing state is the 
antisymmetric combination of winding one states.

Interestingly enough, this is not the whole story. It was found in~\cite{Israel:2007si} 
(see also~\cite{Ashok:2007sf}) that an orientifold described by the parity~(\ref{extpar}) 
alone does not give a consistent crosscap state\footnote{In this paper we will 
use heavily the boundary and crosscap state formalism in boundary conformal
field theory. For an introduction to this formalism we refer the reader to the reviews
\cite{Gaberdiel:2000jr,Brunner:2002em,Brunner:2003zm} and references therein.}  
in the $SL(2)/U(1)$ CFT. The full crosscap state contains a second piece, 
with a wave-function that is localized near the tip of the cigar. The parity 
associated with the localized piece of the crosscap state acts on closed 
string states as 
\begin{equation}
\tilde{\mathcal{P}} = (-)^n \mathcal P = (-)^{F+w} \Omega 
~.
\label{locpar}
\end{equation}
The presence of this parity does not affect the closed string spectrum as
determined with an asymptotic linear dilaton analysis above.\footnote{For 
instance, one can check that the parity \eqref{locpar} does not contribute 
to the Klein bottle amplitude with a term proportional to the volume of the 
asymptotic linear dilaton cylinder. There is only a localized, finite contribution 
from states with zero radial momentum $P$. More details can be found in 
appendix \ref{apploop}, or in ref.\ \cite{Israel:2007si}.} 

The full consistency of the CFT requires that the $\mathcal{N}=2$ Liouville 
potential~(\ref{liouvpot}) is invariant under any parity we want to consider. 
Indeed, one can check that under both parities $\mathcal P$ and 
$\tilde{\mathcal P}$ the $\NN=2$ Liouville potential transforms as follows:
\begin{equation}
\mathcal P, \, \tilde{\mathcal P}\ : \delta \mathcal S_\textsc{l}  (\mu,\bar \mu) 
\longrightarrow \delta \mathcal S_\textsc{l}  (\bar \mu, \mu)\, .
\end{equation}
Consequently, these are symmetries of the CFT at the non-perturbative level if
and only if $\mu \in \mathbb{R}$. 

The fully consistent crosscap state, including both an extensive and a localized 
component that correspond respectively to the parities $\PP$ and $\tilde \PP$, 
can be found with modular boostrap methods, using the results of~\cite{Israel:2007si}. 
It turns out to be similar to the RR part of the supersymmetric 
type II orientifold found in~\cite{Ashok:2007sf}, and reads
\begin{multline}
\label{crosscapstate}
|\mathcal{C}\rangle  = |\mathcal{C};{\rm ext}\rangle + 
|\mathcal{C};{\rm loc}\rangle \\
=\int_{0}^\infty \!\!{\rm d}P\, \sum_{\eta=\pm 1} \left[ \sum_{w \in 2\mathbb{Z}+1}
\Psi_{\rm ext} (P,\tfrac{w}{2},\eta ) |\mathcal{C}; P,\tfrac{w}{2};\eta \rrangle_{R}
+ \sum_{w \in 2\mathbb{Z}}
\Psi_{\rm loc} (P,\tfrac{w}{2},\eta ) |\mathcal{C}; P,\tfrac{w}{2};\eta \rrangle_{R}\right], 
\end{multline}
with the wave-functions\footnote{The wave-functions in 
eqns.~(\ref{waveorienta}, \ref{waveorientb}) are identified with the coefficients of the 
one-point functions on the disc for closed string modes with radial momentum $P$, 
winding $w$ and zero angular momentum $n$, in the RR sector.}
\begin{subequations}
\begin{align}
\label{waveorienta}
\Psi_{\rm ext} (P,\tfrac{w}{2},\eta ) &= \frac{\sqrt{2}}{4\pi^2}\, 
\mu^{iP-\frac{w}{2}}\bar\mu^{iP+\frac{w}{2}}
\frac{\Gamma(1-2iP)\Gamma(-2iP)}{
\Gamma(1-iP +\tfrac{w}{2}) \Gamma (-iP-\tfrac{w}{2})}\,  \\
\label{waveorientb}
\Psi_{\rm loc} (P,\tfrac{w}{2},\eta ) &= \frac{\sqrt{2}}{4\pi^2}\, \eta (-)^{\tfrac{w}{2}} 
\mu^{iP-\frac{w}{2}}\bar\mu^{iP+\frac{w}{2}}
\frac{\Gamma(1-2iP)\Gamma(-2iP)}{
\Gamma(1-iP +\tfrac{w}{2}) \Gamma (-iP-\tfrac{w}{2})} \cosh \pi P \,  .
\end{align}
\end{subequations}
$|\mathcal{C}; P,w/2;\eta \rrangle_{R}$ denote B-type crosscap 
Ishibashi states with superconformal gluing conditions $G^\pm=\eta \tilde G^\pm$ on 
the real axis. The flat space-time part of these Ishibashi states is the standard 
one associated with the parity $(-)^F\Omega$. Although $\mu=\bar\mu$ in our 
case, it will still be useful to keep the holomorphic/anti-holomorphic dependence on
$\mu$ explicit in the crosscap (and later boundary state) wave-functions.

The orientifold we are discussing here is space-filling, hence we will call it an O$'$5-plane. 
Notice that it sources only modes in the RR sector with zero momentum $n$.
The extensive part of the orientifold sources odd winding modes, which are all massive,
whereas the localized part sources even winding modes. The zero winding mode is
a massless RR field. Since there are no massless, non-dynamical RR tadpoles from the 
extensive part of the orientifold, there is no need to add extra D-branes for consistency 
\cite{Israel:2007nj}. Thus, we have a consistent theory of closed strings with a purely 
bosonic spectrum. This should be contrasted with the corresponding
ten-dimensional case \cite{Sagnotti:1995ga,Sagnotti:1996qj}, where D-branes need
to be added to cancel a non-vanishing RR tadpole.

\subsubsection*{A brief comment on the relation with NS5-brane configurations}

There is a configuration of NS5-branes with an orientifold plane 
in ten-dimensional type 0A string theory, whose near-horizon region, 
in a suitable limit, will be described by the type 0B non-critical string theory
on \eqref{noncritback} in the presence of the O$'$5-plane. This configuration 
involves two orthogonal NS5-branes and an O$'$4-plane stretched along the 
following directions:
\bea
NS5~~:~ &&0~1~2~3~4~5~ ~ ~ {\rm at}~x^6=0\,, x^7=x^8=x^9=0
\nonumber\\
NS5'~~:~ &&0~1~2~3~8~9~ ~ ~{\rm at}~x^6=L\,, x^4=x^5=x^7=0
\nonumber\\
O'4~ ~ :~ &&0~1~2~3~6~ ~ ~ ~~\,{\rm at}~ x^4=x^5=x^7=x^8=x^9=0~.
\nonumber
\eea
The O$'$4-plane is the standard O$'$ plane in ten dimensions associated with the
orientifold projection $(-)^F \II_5 \Omega$ ($\II_5$ is the reversal parity in the 
transverse space $x^i \to -x^i$, with $i=4,5,7,8,9$.\footnote{A mirror description 
of this setup is given by wrapped D-branes with an orientifold in the deformed 
conifold geometry as in~\cite{Ooguri:1997ih}.} 

One can argue, as in \cite{Giveon:1999px}, that the tachyon-free type 0B non-critical 
string theory of this section describes the near-horizon geometry of the above 
configuration in a double-scaling limit, where the asymptotic string coupling
$g_s \to 0$ and $\tfrac{L}{\sqrt{\alpha'} g_s}$ is kept fixed. Apparently, as 
we take this near-horizon limit, the bulk tachyon decouples and is left outside 
the near-horizon throat \cite{Israel:2007nj}. One can think of this situation as 
the exact opposite of localized closed string tachyons in non-supersymmetric 
orbifolds~\cite{Adams:2001sv}.

Having said this, we could proceed with the above fivebrane configuration to 
construct a HW setup that realizes the electric description of OQCD
\cite{Armoni:2004uu}. The HW setup requires, besides the fivebranes and 
the O$'$4-plane, $N_c$ D4-branes parallel to the O$'$4-plane suspended 
between the NS5-branes along the 6-direction and $N_f$ D6-branes along 
the 0123789 plane. Then, keeping the bulk tachyon at its unstable maximum, 
we could further use what we know from similar supersymmetric configurations 
in type IIA to argue for Seiberg duality in OQCD and recover the results of section 
\ref{sec:gaugetheory}. We will not follow this route here, instead we will go 
to the non-critical description, which allows for more explicit control in a 
tachyon-free environment and argue for Seiberg duality there.

%%%%%%%%%%%%%%%%%%%%%%%%%%%%%
\subsection{D-branes}
\label{branesect}

$\mathcal{N}=1$ SQCD can be realized in the type IIB non-critical 
background~(\ref{noncritback}) with an appropriate combination of 
localized and extended B-type branes in $SL(2)/U(1)$~\cite{Fotopoulos:2005cn} 
(see also~\cite{Ashok:2005py,Murthy:2006xt}, and~\cite{Eguchi:2003ik} 
for earlier work). The boundary states we will use are the same as those 
of the supersymmetric setup. These ``dyonic'' branes are not elementary in 
oriented type 0B string theory, however they are elementary in the 
presence of the O$'$5 orientifold.

\subsubsection*{Color branes}

The ``color'' branes, $i.e.$ the branes that will provide the gauge degrees of
freedom, are D3-branes in the type 0$'$B theory that are localized near the 
tip of the cigar and have Neumann boundary conditions in $\mathbb{R}^{3,1}$. 
Their open string spectrum is made only of the {\it identity representation} 
of the $\mathcal{N}=2$ superconformal algebra, see app.~\ref{appreps} for 
more details. 

In the $SL(2)/U(1)$ boundary conformal field theory, the corresponding 
boundary states obey B-type boundary conditions and can be expressed 
as a linear combination of B-type boundary Ishibashi states \cite{Fotopoulos:2005cn}
\beq
\label{branesaa}
|\BB;P,\tfrac{w}{2};\eta\rangle\rangle_\textsc{NSNS/RR}~, ~ ~ P\in \R^+~, ~ w \in \Z\, , ~
\eta=\pm 1\, .
\eeq
These Ishibashi states have non-vanishing couplings to winding closed string 
modes only. They are associated to $\mathcal{N}=2$ continuous representations with 
$Q_R = -\tilde{Q}_R = w$ in $SL(2)/U(1)$ and obey in $\R^{3,1}$ the 
standard Neumann boundary conditions. As with the crosscap 
Ishibashi states above, the label $\eta=\pm 1$ refers to the superconformal 
boundary conditions.\footnote{Our definition, which follows standard 
practice, has the property
\begin{equation}
\label{appBaa}
_{NSNS}\langle \eta |e^{-\pi T H_c} | \eta' \rangle_{NSNS} \sim 
\vartheta \left[ {0 \atop \tfrac{1-\eta\eta'}{2} }\right](iT)
~, ~ ~
_{RR}\langle \eta |e^{-\pi T H_c} | \eta' \rangle_{RR} \sim 
\vartheta \left[ {1 \atop \tfrac{1-\eta\eta'}{2} }\right](iT)\, , \nonumber
\end{equation}
where $\vartheta \left[ {a \atop b} \right](\tau)$ are standard theta functions.}

Boundary states with $\eta=\pm 1$ (called respectively ``electric'' or ``magnetic'') 
are separately GSO-invariant in type 0B string theory. The orientifold action, 
however, maps one to the other. More specifically, the action of the parities
(\ref{extpar}, \ref{locpar}) on the Ishibashi states~\eqref{branesaa} is 
\begin{subequations}
\begin{align}
\label{branesab}
\mathcal{P},\, \tilde{\mathcal P}\  |\BB;P,\tfrac{w}{2};\eta\rrangle_{NSNS}
&=(-)^{w+1}|\BB;P,-\tfrac{w}{2};-\eta\rrangle_{NSNS}\, ,
\\
\label{branesac}
\mathcal{P},\, \tilde{\mathcal P}\ |\BB;P,\tfrac{w}{2};\eta\rrangle_\textsc{rr\hphantom{nn}}&
=(-)^{w+1}|\BB;P,-\tfrac{w}{2};-\eta\rrangle_{RR}
\end{align}
\end{subequations}
Then one can check that the D3-brane boundary state, which is invariant
under the both parities $\PP, \tilde \PP$ is of the same form as the boundary state
of the BPS D3-brane in type IIB string theory obtained in~\cite{Fotopoulos:2005cn}
\beq
\label{branesae}
|D3\rangle=\sum_{a=NSNS,RR} \int_0^\infty {\rm d}P\,  \sum_{w \in \Z}
\Phi_{a} \left(P,\tfrac{w}{2}\right) \Big[ |\BB;P,\tfrac{w}{2};+ \rrangle_{a} +(-)^{w+1}
|\BB;P,\tfrac{w}{2};- \rrangle_{a}\Big] \, ,
\eeq
where
\begin{subequations}
\begin{align}
\label{wavecoloura}
\Phi_{{NSNS}}(P,m)&= \frac{\sqrt{2}}{32\pi^2}\,
\mu^{iP-m}\bar\mu^{iP+m}
\frac{\Gamma(\frac{1}{2}+m+iP)\Gamma(\frac{1}{2}-m+iP)}
{\Gamma(2iP)\Gamma(1+2iP)}
\, ,\\
\label{wavecolourb}
\Phi_{RR}(P,m)&= \frac{\sqrt{2}}{32\pi^2}\,
\mu^{iP-m}\bar\mu^{iP+m}
\frac{\Gamma(1+m+iP)\Gamma(-m+iP)}
{\Gamma(2iP)\Gamma(1+2iP)}\, .
\end{align}
\end{subequations}
Notice that this boundary state does not carry any labels, $i.e.$ it has no open 
string modulus.

The annulus and M\"obius strip amplitudes for this brane are presented 
in appendix~\ref{apploop}. The former is identical to the D3-brane annulus 
amplitude in type IIB found in~\cite{Fotopoulos:2005cn}, see 
eq.~(\ref{annuluscol}), and vanishes by supersymmetry. The latter 
contains a contribution from the Ramond sector only, see eq.~(\ref{mobiuscol}), 
hence this amplitude breaks explicitly the Bose-Fermi degeneracy and is
responsible for the supersymmetry breaking in our D-brane setup. Adding the two 
contributions, we can read off the spectrum of massless states  
\begin{multline}
\label{elecab}
\AA_{{\rm D}3-{\rm D}3}+\MM_{D3;{\rm O}'5} = \\= 
V_4 \int_0^\infty\! \frac{\di t}{2t} \frac{1}{(16\pi^2 t)^2}
\left[ \frac{N_c^2}{2} (2+\mathcal{O} (q) )_{\textsc{ns}_+} 
- \frac{N_c(N_c \pm 1)}{2} (2+\mathcal{O} (q) )_{\textsc{r}_+} \right]~.
\end{multline}
The NS sector contains a $U(N_c)$ gauge boson. The R sector contains 
a Dirac fermion transforming in the symmetric (upper sign) or antisymmetric 
(lower sign) representation of $U(N_c)$, depending on the sign of the 
orientifold charge.

\subsection*{Flavor branes}

The ``flavor'' branes are space-filling D5-branes labeled by a continuous 
variable $s \in \mathbb{R}^+$ parametrizing the minimum distance of the brane
from the tip of the cigar. A second continuous parameter $\theta \in [0,2\pi)$ 
parametrizes the value of a Wilson loop around the compact direction of the 
cigar. In the asymptotic cylinder part of the cigar, the D5-branes are 
double-sheeted and look like D-$\bar {\rm D}$ pairs (without an open 
string tachyon however). This geometry is responsible for the 
$U(N_f)\times U(N_f)$ global symmetry of the four-dimensional gauge theory
that we will engineer. Moreover, although space-filling, the flavor D5-branes are not 
charged under a non-dynamical six-form RR potential in six dimensions.

The full D5 boundary states are given in terms of the 
same B-type Ishibashi states as the color branes~(\ref{branesaa})
\begin{equation}
|D5;s,\tfrac{\theta}{2\pi}\rangle=
 \sum_{a=NSNS,RR} \int_0^\infty dP \sum_{w \in \Z}
\Phi_a \left(s,\theta;P,\tfrac{w}{2}\right) \left( |\BB;P,\tfrac{w}{2};+ \rrangle_a 
-(-)^{w}
|\BB;P,\tfrac{w}{2};- \rrangle_a \right)\, ,
\label{branesai}
\end{equation}
where now
\begin{subequations}
\begin{equation}
\label{waveflavoura}
\Phi_{NSNS} (s,\theta;P,\tfrac{w}{2})= \frac{\sqrt{2}}{16\pi^2}\, 
(-)^w e^{-i w \theta}
\mu^{iP-\frac{w}{2}}\bar\mu^{iP+\frac{w}{2}}
\frac{\Gamma(-2iP)\Gamma(1-2iP)}
{\Gamma(\frac{1}{2}-\tfrac{w}{2}-iP)\Gamma(\frac{1}{2}+\tfrac{w}{2}-iP)} \cos(4\pi s P)
\, ,
\end{equation}
\begin{equation}
\label{waveflavourb}
\Phi_{RR}(s,\theta;P,\tfrac{w}{2})= \frac{\sqrt{2}}{16\pi^2}\, 
(-)^w e^{-i w \theta}
\mu^{iP-\frac{w}{2}}\bar\mu^{iP+\frac{w}{2}}
\frac{\Gamma(-2iP)\Gamma(1-2iP)}
{\Gamma(1-\tfrac{w}{2}-iP)\Gamma(\tfrac{w}{2}-iP)} \cos(4\pi s P)
\, .
\end{equation}
\end{subequations}

The annulus amplitude for open strings stretched between identical flavor 
branes ({\it 5-5 strings}) is given by eq.~(\ref{annulusflav}) in 
appendix~\ref{apploop}. The massless spectrum comprises of
an ${\mathcal N}=1$ chiral multiplet (with the quantum numbers of a massless 
meson), which is part of a continuous spectrum of modes.
Vacuum expectations values of the scalar fields in this
multiplet should be interpreted as parameters of the four-dimensional gauge 
theory~\cite{Fotopoulos:2005cn}. The massive spectrum contains a vector 
multiplet, which accounts for the gauge degrees of freedom on the D5-brane. 
The positive mass shift in this multiplet is due to the linear dilaton.

The action of the orientifold parity on the open strings attached to the flavor branes
can be read from the corresponding M\"obius strip amplitudes, which appear in 
appendix \ref{apploop}. As before, the M\"obius strip amplitudes are non-zero only
in the Ramond sector, hence they leave the bosons unaffected but project the fermions.
In particular, the fermionic superpartner of the massless chiral multiplet
transforms, after the orientifold projection, in the (anti)symmetric representation of 
the diagonal $U(N_f)$ flavor group (the opposite projection compared to that
of the gauginos). 

The non-vanishing M\"obius strip amplitude manifestly shows that 
spacetime supersymmetry is broken on the flavor branes and leads to 
a net force between the flavor D5-branes and the orientifold plane, which 
we discuss in appendix \ref{appMob}. This force, which arises as a one-loop 
effect on the flavor branes, has no consequences on the dynamics of the 
four-dimensional OQCD theory that will be engineered in the next subsection. 
Indeed, we will consider the open string dynamics in a low energy decoupling 
limit where the dynamics on the flavor branes are frozen. In this limit, only fields
from the {\it 3-3} and {\it 3-5} sectors are allowed to run in loops.

\subsection*{Flavor open strings}

Open string stretched between flavor branes and color branes ({\it 3-5 strings}) 
transform in the fundamental representation of $U(N_c)$. From the relevant 
annulus amplitude, see eq.~(\ref{annul35}), one gets a non-tachyonic, 
supersymmetric open string spectrum either from a flavor brane with $\theta=0$ 
($|D5(s,0)\rangle$) or a flavor anti-brane $\overline{|D5(s,\nicefrac{1}{2})\rangle}$ 
with $\theta=\pi$ \cite{Fotopoulos:2005cn}. For these branes and $N_c$ color 
branes the {\it 3-5} sector includes the following light degrees of freedom
\begin{itemize}
\item A flavor brane with $\theta=0$ gives an ${\mathcal N}=1$ massive 
vector multiplet with four-dimensional mass $m=\sqrt{2s^2+1/2}$ in the 
fundamental of $U(N_c)$.
\item A flavor anti-brane with $\theta=\pi$ gives two quark chiral multiplets 
with mass $m=\sqrt{2}s$, in the fundamental and antifundamental of $U(N_c)$. 
\end{itemize}
Only the second case will be relevant for engineering the electric OQCD theory.

%%%%%%%%%%%%%%%%%%%%%%%%%%%
\subsection{Realizing the electric theory}
\label{electricsetup}

We are now in position to reveal the final picture. The electric version 
of OQCD can be obtained in non-critical type 0$'$B string theory 
as the low-energy description of the open string dynamics on 
$N_c$ color branes $|D3\rangle$ and $N_f$ flavor branes of the type 
$\overline{|D5(s,\nicefrac{1}{2})\rangle}$. In this configuration, the 
four-dimensional low-energy degrees of freedom are
\begin{itemize}
\item A $U(N_c)$ gauge field $A_\mu$ from {\it 3-3 strings}.
\item A Dirac fermion in the symmetric or antisymmetric $U(N_c)$ 
representation from {\it 3-3 strings}. Positive orientifold charge gives 
the symmetric representation and negative charge the antisymmetric. 
\item $N_f$ quark multiplets $\Phi$ in the $U(N_c)$ fundamental 
representation and $N_f$ anti-fundamental multiplets $\tilde{\Phi}$ 
from {\it 3-5 strings}. The mass of these multiplets is proportional to 
$s$, the parameter that appears in the labeling of the D5-brane 
boundary state.
\end{itemize}
In addition, from {\it 5-5 strings} we get a massless chiral multiplet 
propagating in five dimensions. From the four-dimensional point of 
view the dynamics of this multiplet are frozen; the vacuum expectation 
value of the scalar component of this multiplet gives a mass to the 
quarks, $i.e.$ it is directly related to the parameter $s$ above 
\cite{Fotopoulos:2005cn}. 

So far we have discussed the low energy open string spectrum, but we 
have not determined unambiguously all the couplings in the low energy 
string field theory Lagrangian.\footnote{From the gauge theory point of view 
this is the UV Lagrangian defined at some scale below the string scale.}
The symmetries of the D-brane setup suggest that the Lagrangian 
includes the usual minimal couplings of the OQCD theory, but, in 
principle, it is possible that additional couplings allowed by the 
symmetries are also present. One could check this by computing 
tree-level correlation functions in open string theory. This is a  
complicated exercise that requires explicit knowledge of open
string correlation functions in the boundary conformal field theory 
of $SL(2)/U(1)$. This information goes beyond the currently available
technology, so we will not pursue such a task any further here. 

Nevertheless, it was pointed out in \cite{Murthy:2006xt} (in a type IIB
context, using the results of a computation in \cite{Ashok:2005py}) that the 
leading asymptotic backreaction of $N_c$ $|D3\rangle$ and 
$N_f$ $\overline{|D5(s,\nicefrac{1}{2})\rangle}$ boundary states
on the profile of the dilaton and graviton fields in the bulk is 
proportional to $N_f-2N_c$. One would expect the factor 
$N_f-3N_c$ from a D-brane setup realizing just $\NN=1$ 
SQCD without any extra couplings. On these grounds, 
ref.\ \cite{Murthy:2006xt} proposed that the gauge theory on
the above D-brane setup is in fact $\NN=1$ SQCD with an
extra quartic superpotential coupling of the form
\beq
\label{quartica}
\int {\rm d}^2\theta \, W=h \int {\rm d}^2\theta \, (Q\tilde Q) (Q\tilde Q)
~
\eeq
where $Q$, $\tilde Q$ are the quark multiplets of $\NN=1$ SQCD
in the fundamental and antifundamental respectively. This 
proposal is consistent with the one-loop beta function of this
theory \cite{Strassler:2005qs}, which is also proportional to $N_f-2N_c$.
In the context of holography, this quartic coupling has been discussed
recently in \cite{Casero:2006pt,Casero:2007jj}.

These observations carry over naturally to our non-supersymmetric case.
Hence, we propose that the above D-brane setup in type 0$'$B string theory 
realizes the electric OQCD-S theory (see table~\ref{tabelectric}) if the orientifold 
charge is positive, or the OQCD-AS theory if the charge is negative, with an 
extra quartic coupling. We will find further evidence for the quartic coupling 
in a moment.

%%%%%%%%%%%%%%%%%%%%%%%%%%%%%%% 
\subsection{D-brane monodromies}

Now we want to examine the behavior of the D3, D5 boundary states 
and the O$'$5 crosscap state under the $\Z_2$ transformation of the 
closed string modulus $\mu$ (the Liouville coupling constant, see 
eq.~\eqref{liouvpot})
\beq
\label{monodraa}
\mu \to -\mu
~.
\eeq
Henceforth, we will refer to this transformation as the $\mu$-transition 
following~\cite{Hori:2005bk}. It has been argued in a similar type IIB 
setting \cite{Murthy:2006xt,Ashok:2007sf} that one can use the 
$\mu$-transition to interpolate between D-brane configurations that 
realize the electric and magnetic descriptions of SQCD with gauge 
groups $SU$, or $SO/Sp$. This fits very nicely with the HW description, 
where $\mu$ corresponds to the separation of the NS5 and NS5$'$ 
branes in the $(x^6,x^7)$ plane.

Note that in the supersymmetric $SU$ case, $\mu$ takes any value 
in the complex plane, so this interpolation can be made continuously 
without ever passing through the strong coupling point at $\mu=0$. 
This should be contrasted with the $SO/Sp$ cases \cite{Ashok:2007sf} 
where $\mu$ is real, so a continuous interpolation from $\mu$ to $-\mu$ 
entails passing through the strong coupling point at the origin at which 
we lose perturbative control over the theory. In the present case of type 
0$'$B string theory we face a similar difficulty, since $\mu$ is again real. 
Our attitude towards this will be the same as in~\cite{Ashok:2007sf}: we 
will perform the $\mu$ transition~\eqref{monodraa} as a $\Z_2$ 
transformation and will account for any missing RR charge at the end 
of this process by adding the appropriate number of D-branes needed 
by charge conservation. Here we make the underlying assumption that 
nothing dramatic can happen along the $\mu$ real line that could affect 
the IR dynamics of the gauge theory on the branes. We discuss this point 
further in the next section.  

To distinguish the boundary and crosscap states of the $\mu$ and $-\mu$ 
theories we will explicitly label them by $\mu$. To implement Seiberg 
duality, we supplement the $\mu$-transition \eqref{monodraa} with the 
following boundary and crosscap state transformations\footnote{Notice 
that we set the D5-brane parameter $s$ to zero. In other words,
we discuss what happens when the quark multiplets in the electric description
are massless.}
\begin{subequations}
\begin{align}
\label{monodrab}
& |D3;\mu \rangle \to |D3;-\mu\rangle~, 
\\
\label{monodrac}
& \overline{|D5;0,\nicefrac{1}{2};\mu\rangle}\to
\overline{|D5;\nicefrac{i}{2},0;-\mu\rangle}=
|D5;0,\nicefrac{1}{2};-\mu\rangle+\overline{|D3;-\mu\rangle}
~,
\\
\label{monodrad}
& |{\rm O}'5;\mu\rangle \to \overline{|{\rm O}'5;-\mu\rangle}
~.
\end{align}
\end{subequations}
These monodromies are different from the ones proposed in 
\cite{Murthy:2006xt,Ashok:2007sf}, so we will take a moment here to explain
each one of them in detail.\footnote{Some of the monodromies proposed in 
\cite{Murthy:2006xt,Ashok:2007sf} do not follow from the $\mu \to -\mu$
transformation and lead to D3-branes with negative tension. They also 
lead to a gauge theory spectrum with unexplained features that does not fit
with the expected Seiberg duality. The monodromies we present here
do not have these problems. Although we discuss a setup realizing 
OQCD, our analysis has analogous implications for the D-brane setups 
that realize $\NN=1$ SQCD with $SU$ or $SO/Sp$ gauge groups, 
which were the subject of \cite{Murthy:2006xt,Ashok:2007sf}.}

First of all, the $\mu \to -\mu$ transformation is equivalent, as a transformation 
of the $\NN=2$ Liouville interaction~\eqref{liouvpot}, to a half-period shift 
$s_{\tilde{\textsc{x}}}$ along the angular direction in winding space, 
$i.e.$ $X_L-X_R \to X_L-X_R +\sqrt{2} \pi$. Hence, the $\Z_2$ transformation
\eqref{monodraa} is equivalent to the multiplication of the D-brane wave-functions 
with the phase $(-)^w$, which follows from the action 
\begin{equation}
s_{\tilde{\textsc{x}}} |\mathcal{B};P,\tfrac{w}{2};\eta \rrangle = 
(-)^w|\mathcal{B};P,\tfrac{w}{2};\eta 
\rrangle \ , \quad s_{\tilde{\textsc{x}}} |\mathcal{C};P,\tfrac{w}{2};\eta \rrangle 
= (-)^w|\mathcal{C};P,\tfrac{w}{2};\eta \rrangle \, .
\label{Ishiact}
\end{equation}
This is consistent with the $\mu$ dependence of the 
wave-functions (\ref{wavecoloura}, \ref{wavecolourb}), (\ref{waveflavoura}, 
\ref{waveflavourb}). 
The first line~\eqref{monodrab} exhibits the transformation of the D3-brane 
($i.e.$ the color brane) according to these rules.

The second line~\eqref{monodrac} presents the transformation of
the D5 boundary states. In order to obtain Seiberg duality, we want 
a process that affects only the physics near the tip of the cigar. In 
particular, we want to keep the value of the Wilson line on the 
D5-branes, as measured in the asymptotic region $\rho \to \infty$, 
fixed during the $\mu$-transition. Otherwise, we simply rotate the whole
configuration by half a period in the angular (winding) direction of the 
cigar to get back the electric description of the gauge theory. 

In order to achieve this, we have to shift the D5 boundary state 
modulus $M=\frac{\theta}{2\pi}=\frac{1}{2}$ to $M=0$ (notice that this offsets the 
effect of the $\mu \to -\mu$ transformation on the wave-functions 
(\ref{waveflavoura}, \ref{waveflavourb})). At the same time, we want 
to keep the brane chiral, $i.e.$ maintain the property $J=M$ of the 
D-brane labels,
where, by definition, $J=\frac{1}{2}+is$. This is important for the 
following reason. The worldsheet action for an open string ending 
on the D5 boundary state is captured by the boundary action
\cite{Hosomichi:2004ph}
\bea
\label{hosoaction}
\SS_{\rm bdy}=&&\oint {\rm d}x \Big[ \bar \lambda \d_x \lambda
-\mu_B \lambda \psi^+ e^{-\frac{\rho_{\rm L}+i X_{\rm L}}{\sqrt 2}}
-\bar \mu_B e^{-\frac{\rho_{\rm L}+i X_{\rm L}}{\sqrt 2}} \psi^- \bar\lambda
\\
&&-\tilde \mu_B(\lambda \bar \lambda-\bar \lambda \lambda)(\psi^+\psi^- - 
i\sqrt 2 \d X)e^{-\sqrt 2 \rho_{\rm L}} \Big]
\, , \nonumber
\eea
where $\lambda$, $\bar \lambda$ are boundary fermions and
$\mu_B$, $\bar \mu_B$, $\tilde\mu_B$ are boundary Liouville couplings
that are related to the $(J,M)$ labels of the branes and the bulk Liouville 
coupling $\mu$ by the following equations\footnote{As in the bulk, see 
eq.~(\ref{renorm}), these relations need to be renormalized in the limit 
$k \to 1^+$~\cite{Murthy:2006xt}.}
\begin{subequations}
\begin{align}
\label{hosocouplingsa}
\tilde \mu_B^{{\rm ren};k=1}& =
\frac{1}{2\pi }|\mu^{{\rm ren};k=1}|\, \cos \pi (2J-1)  ~,
\\
\label{hosocouplingsb}
\mu_B^{{\rm ren};k=1} \bar \mu_B^{{\rm ren};k=1}
&=\frac{2}{\pi}\, |\mu^{{\rm ren};k=1}\,| \sin \pi (J-M)\, \sin \pi (J+M) \, .
\end{align}
\end{subequations}
For $J=M=\frac{1}{2}$ the boundary couplings  $\mu_B$ and $\bar \mu_B$ 
vanish. Demanding that they still vanish after the transformation $M=\frac{1}{2}
\to M=0$ requires that we also set $J=0$, $i.e.$ that we perform
an additional $\Z_2$ transformation on the D5 boundary state
$\tilde \mu_B \to -\tilde \mu_B$.

This explains the transformation appearing in \eqref{monodrac}. 
The equality that expresses the $J=M=0$ boundary state as a 
linear combination of the fundamental D5-brane with $J=M=\frac{1}{2}$
and the D3-brane follows from the character identity \eqref{appAba} in 
appendix \ref{appreps}. The importance of this relation for this setup was 
also pointed out in \cite{Murthy:2006xt} (the character identity \eqref{appAba} 
was observed earlier in \cite{Eguchi:2003ik}). In the HW context, it expresses 
the possibility of the flavor branes to break onto the NS5-brane after the rotation; 
a similar phenomenon was touched upon in~\cite{Israel:2005fn} for 
$\mathcal{N}=2$ setups. 

Finally, in the last line \eqref{monodrad} the bar follows from the 
requirement that we get the same projection on the gauge theory fermions
before and after the $\mu$-transition \cite{Ashok:2007sf}. From the 
$\mu$-transition action on the Ishibashi states, eq.~(\ref{Ishiact}), 
and the expression of the crosscap state~(\ref{crosscapstate}), one observes 
that this transformation is such that the extensive part of the orientifold 
is invariant, while the localized part is reversed. As with the D5-brane, this 
is a consequence of the requirement to have a ``local'' operation, leaving  
the asymptotics invariant. 

Consequently, if we start with the electric description of OQCD that arises in 
a setup that includes the states
\beq
\label{monodrae}
N_c ~ |D3;\mu\rangle~, ~ ~ N_f ~ \overline{|D5;0,\nicefrac{1}{2};\mu\rangle}~, ~ ~ 
\pm |{\rm O}'5 ;\mu\rangle 
\eeq
we end up naively after the $\mu$-transition and the annihilation of $N_c$ 
D3 brane-antibrane pairs with a configuration that comprises of the   
boundary states $(N_f-N_c) \overline{|D3;-\mu \rangle}$ and 
$N_f |D5;0,\nicefrac{1}{2};-\mu \rangle$, together with the crosscap 
$\pm \overline{|{\rm O}'5;-\mu\rangle}$. 

This, however, cannot be the final answer. Notice that in this 
process the $C_0$ RR scalar charge has changed. In appropriate 
normalizations where the $|D3;\mu\rangle$ boundary state 
has unit RR charge, the total charge of the electric configuration is 
\cite{Murthy:2006xt,Ashok:2007sf} (see also appendix~\ref{app:RRcha} 
for a detailed determination of the RR charges)
\beq
\label{monodrag}
Q_e=N_c-\frac{N_f}{2}\pm 2
~.
\eeq
The $\pm$ sign in this relation is the same as the $\pm$ sign 
appearing in front of the crosscap state in \eqref{monodrae}. We 
remind the reader that the $+$ sign corresponds to gauginos in 
the symmetric representation and the $-$ sign to gauginos in the 
antisymmetric.

The RR scalar charge of the $-\mu$ configuration is equal to
\beq
\label{monodrai}
Q_m=N_c-\frac{N_f}{2}\mp 2
~.
\eeq
Therefore, charge conservation requires the creation of new charge during
the $\mu$-transition. This is a familiar effect in HW setups with orientifolds
\cite{Evans:1997hk}. Remember that massless RR charge is carried only by 
the $|\CC;{\rm loc}\rangle$ part of the $|{\rm O}'5\rangle$ crosscap state (see
eq.~\eqref{crosscapstate}). Hence, it is natural to proclaim that the new 
charge is carried by the only other localized object in the game, the 
D3-branes. Adding four D3 or $\overline {{\rm D}3}$ boundary states 
in the $-\mu$ configuration, we are canceling the deficiency (or surplus) of four 
units of scalar charge to obtain the following magnetic configuration
\beq
\label{monodraj}
(N_f-N_c\mp 4) \overline{|D3;-\mu \rangle} ~, ~ ~ 
N_f |D5;0,\nicefrac{1}{2};-\mu \rangle~, ~ ~ 
\pm \overline{|{\rm O}'5;-\mu\rangle}
~.
\eeq
The low-energy open string degrees of freedom of this configuration 
are string theory's ``prediction'' for the magnetic version of OQCD.

%%%%%%%%%%%%%%%%%%%%%%%%%%%
\subsection{The magnetic theory}
\label{magneticsetup}

Low-energy open string degrees of freedom arise in three different sectors
in the D-brane configuration \eqref{monodraj}: {\it 3-3 strings, 3-5 strings} and
{\it 5-5 strings}. These can be determined from the annulus and M\"obius strip
amplitudes that appear in appendix \ref{apploop}. They are in fact the same
as those appearing in subsection \ref{electricsetup} for the electric setup
of OQCD. The main difference is the all important change in the rank
of the gauge group from $N_c$ to $N_f-N_c\mp 4$. Besides that, the 
$\mu$-transition exhibits a self-duality between the electric and magnetic
setups. Indeed, both setups involve the same boundary and crosscap states
(modulo a sign change in the RR sector which is naturally accounted for
by the fact that we have rotated the theory in the bulk along the angular direction
of the cigar).

The self-duality of the configuration matches very nicely with the proposal
in subsection~\ref{electricsetup} that the low-energy theory on the D-branes 
is OQCD with a quartic coupling. In the large $N$ limit, the theory
is planar equivalent to $\NN=1$ SQCD with the quartic coupling \eqref{quartica}.
Seiberg duality for $\NN=1$ SQCD with a quartic coupling gives back the
same theory with an inverse coupling constant for the quartic interaction 
and therefore exhibits the same self-duality property~\cite{Strassler:2005qs}. 
The dual Lagrangian can be obtained from the magnetic theory with a 
supersymmetric mass term for the magnetic meson if we integrate out the 
magnetic meson. For $N_f \neq 2N_c$, one recovers in the far infrared the 
usual duality between the electric and magnetic descriptions of SQCD 
without quartic couplings. The case $N_f=2N_c$ is special. In this case, 
the quartic coupling becomes exactly marginal in the infrared. 

The string theory picture of this section, suggests a similar state of affairs 
also in OQCD. This is certainly true in the large $N$ limit, because 
of planar equivalence, but we would like to propose here that this picture 
extends also at finite $N$. Furthermore, we consider this picture as evidence 
for the validity of the finite $N$ Seiberg duality of section \ref{sec:gaugetheory} 
in the absence of quartic couplings.

%%%%%%%%%%%%%%%%%%%%%%%%%%%%%%%%%%%%%%%%%%%%%%
\section{Evidence for the duality}

The string theory analysis of the previous section motivates an electric-magnetic
duality for OQCD as postulated in section \ref{sec:gaugetheory}. This duality
shares many common features with Seiberg duality in $\NN=1$ SQCD in the 
$SU$ and $SO/Sp$ cases. However, unlike the supersymmetric case, here 
we cannot use, in principle, the arsenal of supersymmetry to perform a set 
of non-trivial consistency checks. For instance, we cannot use holomorphy to fix 
the exact quantum form of potential terms, there is no known exact beta function 
formula \cite{Novikov:1983uc,Shifman:1986zi,Shifman:1991dz} (except at large 
$N$ \cite{Armoni:2003gp}) and there is no superconformal symmetry or chiral 
ring structure that fixes the IR features of a class of special operators. What 
support can we then find in favor of our claim given that OQCD is a 
non-supersymmetric gauge theory? In this section, we would like to summarize 
the evidence for the proposed duality and discuss its viability as a conjecture 
in non-supersymmetric gauge theory.

%%%%%%%%%%%%%%%%%%%%%%%%%%
\subsection{Duality at large $N$}

First of all, there is a sense in OQCD, in which we can expand around 
a supersymmetric point. At infinite $N$, namely when both 
$N_c \rightarrow \infty$ and $N_f \rightarrow \infty$, with 
$g_{\rm YM}^2 N_c$ and $N_f/N_c$ kept fixed, the dynamics of the 
electric and magnetic theories is almost identical to the dynamics 
of $\NN=1$ SQCD. The reason is, as we mentioned above, 
``planar equivalence'' \cite{Armoni:2004uu}: the planar 
non-supersymmetric electric (magnetic) theory is non-perturbatively 
equivalent to the supersymmetric electric (magnetic) theory in 
the common sector of C-parity even states \cite{Unsal:2006pj}. 
This argument, by itself, is sufficient for the duality to hold.

A consistency check of the proposed duality at infinite $N$ is a 
one-loop calculation of the Coleman-Weinberg potential for the 
``squark'' field (this will be discussed in more detail in the next 
section). The potential remains flat. This is consistent with the 
existence of a moduli space in SQCD.

In addition, both the magnetic and the electric non-supersymmetric 
theories admit an NSVZ beta function \cite{Armoni:2003gp} at large 
$N$. The large-$N$ NSVZ beta function supports the existence of a 
conformal window as in Seiberg's original paper \cite{Seiberg:1994pq}.

We wish to add that the infinite-$N$ equivalence with the supersymmetric 
theory does not involve any fine-tuning. Once the tree level Lagrangian 
is given, quantum corrections will preserve the ratio between couplings 
as if the theory was supersymmetric. 

We conclude that planar equivalence with SQCD is a non-trivial and 
non-generic statement that proves the proposed duality at infinite $N$. 
Now the question about the fate of the duality becomes a question about 
the effects of $1/N$ corrections. Our evidence for a finite-$N$ duality is 
weaker. One argument is that if a certain projection (``orientifolding'') is made 
on both the electric and magnetic original theories, the IR-duality should still 
hold. Additional arguments can be made on the basis of anomaly matching 
and the string realization.

%%%%%%%%%%%%%%%%%%%%%%%%%%%%%%
\subsection{Anomaly matching}

A non-trivial check of consistency that we can always perform independent 
of supersymmetry, is 't Hooft anomaly matching, $i.e.$ we should verify that 
global anomalies are the same in the UV and IR descriptions of the theory. 
For concreteness, let us consider here the case of OQCD-AS 
(similar statements apply also to OQCD-S). The global $SU(N_f)^3$, 
$SU(N_f)^2 U(1)_R$, $U(1)_R$ and $U(1)^3_R$ anomalies are summarized 
for the electric and proposed magnetic descriptions of OQCD-AS in table 
\ref{tab:anomaly}. As before, we use the notation $\tilde{N_c}=N_f-N_c+4$
and the terms in each box are ordered as (``gluino'') + (quarks) in the electric 
theory and (``gluino'') + (quarks) + (``mesino'') in the magnetic theory. 
$d^2(R)\delta^{ab}$ and $d^3(R)d^{abc}$ for the representation $R$ are 
respectively the traces $\tr_R \, T^a T^b$, $\tr_R \, T^a \{ T^b, T^c\}$.\footnote{By 
definition $d^{abc}=\tr_{\Yfund} \, T^a \{ T^b, T^c\}$.} In table \ref{tab:anomaly} 
we make use of the following relations:
\beq
\label{eviaa}
d^2(\Ysymm)=(N_f+2)d^2(\operatorname{\Yfund})~, ~ ~ 
d^3(\operatorname{\Ysymm})=(N_f+4)d^3(\operatorname{\Yfund})
~.
\eeq

It is worth noticing that up to factors of $2$ the matching works precisely as 
the matching of the anomalies in the supersymmetric $SO(N)$ case. This is 
not surprising, since the fermions in our model carry the same dimensions 
(up to factors of 2) as the fermions in the $SO(N)$ models.

The perfect matching of the above anomalies, including $1/N$ corrections, is 
our first independent consistency check for the duality directly in gauge theory. 
It suggests that the duality, as we formulated it in section \ref{sec:gaugetheory}, 
may hold even at finite $N$. There are known non-supersymmetric cases, however, 
where 't Hooft anomaly matching is misleading as a basis for the proposal of an 
IR effective theory \cite{Terning:1997xy,Brodie:1998vv}. These are cases where 
the matching works in a non-trivial manner, but one can still argue that the 
so-proposed IR effective theory cannot be the correct one. Our case is different,
however, because of planar equivalence at infinite $N$, which fixes the
basic structure of the duality.

\begin{table}[!t]
\begin{center}
\begin{tabular}{|l|l|l|}
\hline
 & Electric & Magnetic \\
\hline \hline
& & \\
$\SU(N_f)^3$ & 
$0 + N_cd^3(\Yfund)$ & $0 + \tilde{N_c}(-d^3(\Yfund))+d^3(\Ysymm)$\\
& $=N_cd^3(\Yfund)$ & $=N_cd^3(\Yfund)$\\
& & \\
\hline \hline
& & \\
$\SU(N_f)^2\U(1)_R$ &
$0 + N_c\left(\frac{-N_c+2}{N_f}\right)d^2(\Yfund)$ & $0 
+ \tilde{N_c}\left(\frac{N_c-N_f-2}{N_f}\right)d^2(\Yfund)+$\\
& $=\frac{-N_c^2+2N_c}{N_f}d^2(\Yfund)$ 
& $+\left(\frac{N_f-2N_c+4}{N_f}\right)d^2(\Ysymm)=\frac{-N_c^2+2N_c}{N_f}d^2(\Yfund)$\\
& & \\
\hline \hline
& & \\
$\U(1)_R$ &
$(N_c^2-N_c)+$ 
& $(\tilde{N_c^2}-\tilde{N_c})+2\left(\tilde{N_c}N_f\frac{N_c-N_f-2}{N_f}\right)$+\\
& $+2\left(N_cN_f\frac{-Nc+2}{N_f}\right)$
& $2\left(\tfrac{1}{2}(N_f^2+N_f)\frac{N_f-2N_c+4}{N_f}\right)$\\
& $=-N_c^2+3N_c$ & $=-N_c^2+3N_c$\\
& & \\
\hline \hline
& & \\
$\U(1)_R^3$ &
$(N_c^2-N_c)+$ 
& $(\tilde{N_c^2}-\tilde{N_c})+2\left[\tilde{N_c}N_f\left(\frac{N_c-N_f-2}{N_f}\right)^3\right]+$ \\
& $+2\left[N_cN_f\left(\frac{-Nc+2}{N_f}\right)^3\right]$
& $2\left[\tfrac{1}{2}(N_f^2+N_f)\left(\frac{N_f-2N_c+4}{N_f}\right)^3\right]$\\
& $=N_c\left(N_c-1-2\frac{(N_c-2)^3}{N_f^2}\right)$ 
& $=N_c\left(N_c-1-2\frac{(N_c-2)^3}{N_f^2}\right)$\\
& & \\
\hline
\end{tabular}
\caption{\it 't Hooft anomaly matching for OQCD-AS.}
\label{tab:anomaly}
\end{center}
\end{table}

%%%%%%%%%%%%%%%%%%%%%%%%%%%%%%%%%
\subsection{Hints from the string theory realization}

The string theory realization of section \ref{sec:embedding} cannot be taken
as a proof for the proposed duality, but we believe it gives useful hints. 
Let us outline some of the pertinent issues.

What we presented in section \ref{sec:embedding} is a procedure -- 
the $\mu$-transition -- that relates two distinct configurations of 
D-branes and O-planes (exactly described in a non-trivial underlying 
two-dimensional worldsheet CFT via the appropriate boundary and 
crosscap states). This procedure as a continuous process involves a 
number of non-trivial steps. Starting from the electric description with 
$\mu=\mu_0 \in \R^+$ we slide $\mu$ along the real axis until it goes 
through zero to $-\mu_0$. At zero $\mu$ the background string theory 
develops a linear dilaton singularity and becomes strongly coupled. 
This is the first non-trivial effect. Then as we emerge from zero $\mu$ 
four extra D-branes have been created and $N_c$ D-$\bar{\rm D}$ pairs 
annihilate via open string tachyon condensation.

In order to argue for Seiberg duality in this setup we have to show that 
the IR dynamics on the branes are unaffected by the above procedure. 
For instance, one should show that any open string process that contributes 
in the extreme IR is independent of the sign of $\mu$. This is plausible, given 
the fact that the boundary states are the same as in the corresponding 
$SO$/$Sp$ setup in type IIB string theory \cite{Ashok:2007sf} and the crosscap 
state is the RR part of its supersymmetric counterpart. We have not been able 
to show this explicitly however.

Another source of intuition that the $\mu$-transition produces an actual 
Seiberg dual is based on the fact that the closed string modulus $\mu$ 
controls the bare coupling $g_{\rm YM}$ on the branes
\beq
\label{eviba}
g_\textsc{YM}^2\sim \nicefrac{1}{|\mu|}
~.
\eeq
Hence, changing $\mu$ does not affect the IR dynamics. This naive picture 
could be spoiled if, for example, some extra non-trivial strong coupling IR dynamics 
take place at $\mu=0$, so that when we emerge on the other side of the $\mu$-line 
the IR dynamics on the branes are not anymore related with the IR dynamics 
on the brane setup of the electric description. In the supersymmetric case without 
orientifolds, type IIB string theory on $\R^{3,1}\times SL(2)_1/U(1)$ is related 
to type IIB string theory on the resolved conifold in a double scaling limit 
\cite{Ooguri:1995wj,Giveon:1999px}. In this limit, $g_s$ and the volume of the 
blown-up $S^2$ cycle are taken to zero in such a way that D5-branes wrapping the 
$S^2$ have a finite tension \cite{Giveon:1999px}. The tension vanishes when $\mu$ 
on the $SL(2)/U(1)$ side becomes zero. In that case, the linear dilaton 
singularity is a signal of the corresponding conifold singularity. It is known 
\cite{Strominger:1995cz,Becker:1995kb,Ooguri:1996me,Saueressig:2007dr} that 
the singularity of a slightly resolved conifold can be explained by taking into
account the instanton contributions associated to D1-branes wrapping the 
$S^2$. These contributions are already present in the gauge theory on the
D-brane setup of the electric description, so no abrupt change is anticipated 
as we go through the singularity. 

The situation is more involved in the presence of an O4-plane. The 
NS5-branes act as domain walls for the orientifold charge, so when an 
O4-plane goes through an NS5-brane its charge changes (we can see this 
effect clearly in the non-critical description: the culprit for the extra
charge is the localized part of the crosscap state $|\CC;{\rm loc}\rangle$ -- 
see eq.\ \eqref{crosscapstate}). A consequence of this is a shift in RR charge
as we go through the $\mu$-transition, which has to be compensated by the
creation of 4 (anti)D-branes. This a clear signal of a phase transition that
occurs at $\mu=0$. Still, what we know about Seiberg duality in SQCD with
$SO/Sp$ gauge groups suggests that there are no strong coupling dynamics 
that affect the IR physics during this phase transition. Given the similarities of the 
supersymmetric setup with the setup realizing OQCD, we are tempted to 
speculate that an analogous conclusion can be reached in the absence 
of supersymmetry, although, clearly, it would be desirable to substantiate 
this conclusion with a stronger argument.

%%%%%%%%%%%%%%%%%%%%%%%%%%%%%%%%%%%%%%%%%%%%%%
\section{Implications in gauge theory}

\noindent
In the previous section we provided evidence for a Seiberg duality
between two non-supersymmetric (OQCD) gauge theories. 
For the moment let us accept this duality as a correct statement 
and discuss its implications. 

At infinite $N$ the dynamics of the electric and magnetic theories 
is supersymmetric in the sector of C-parity even states due 
to planar equivalence \cite{Armoni:2004uu}. 
The immediate implications of this statement are: a quantum moduli space, 
a conformal window at ${3 \over 2} N_c \leqslant N_f \leqslant 3N_c$ and 
confinement at $N_c \leqslant N_f < {3 \over 2} N_c$. For 
$N_c+1<N_f<\frac{3}{2}N_c$ the magnetic description is IR free and 
provides a good effective description of the IR dynamics.

At finite $N$ the non-supersymmetric effects become more significant and 
determining exactly the IR dynamics becomes a complicated problem. As 
in ordinary QCD, a Banks-Zaks analysis of the two-loop beta function reveals 
that there should be a range of $N_f$ where the IR dynamics are captured by
a scale-invariant theory of interacting quarks and gluons. One of the most 
interesting implications of our proposal for a Seiberg duality in OQCD is a 
precise prediction for the exact range of the conformal window. 

The one-loop beta function coefficient of the electric OQCD-AS theory is 
$\beta = 3N_c - N_f + {4 \over 3}$. Similarly, the one-loop beta function 
coefficient of the magnetic theory is $\beta = 3(N_f-N_c +4) - N_f +{4 \over 3}$. 
Since the upper and lower parts of the conformal window are determined by 
the vanishing of the one-loop beta function coefficients of the electric and 
magnetic theories, we expect a conformal window for OQCD-AS when
\beq
{3\over 2} N_c - {20 \over 3}  \leqslant  N_f \leqslant 3N_c+ {4 \over 3} 
\,\,\,\, , \,\,\,\, N_c > 5 \, .
\eeq   
When the ``gluino'' is in the symmetric representation (the OQCD-S theory) 
we expect a conformal window when
\beq
{3\over 2} N_c + {20 \over 3}  \leqslant  N_f \leqslant 3N_c -  {4 \over 3} 
\,\,\,\, , \,\,\,\, N_c> 5
\, .
\eeq   
The restriction on the number of colors $N_c > 5$ is explained below. 

\begin{figure}[t!]
\centerline{\includegraphics[width=14cm]{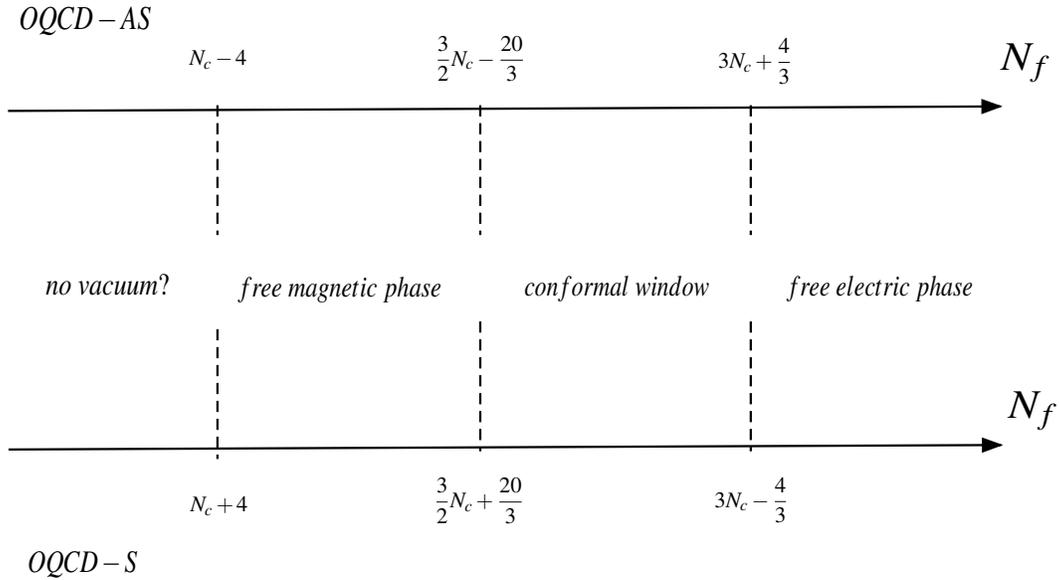}}
\caption{\it The conjectured phase structure of OQCD as a function 
of the number of flavors $N_f$. At infinite $N$ we recover the phase structure of
$\NN=1$ SQCD with gauge group $U(N_c)$. This picture makes sense for $N_c>5$.}
\label{fig:OQCDphases}
\end{figure}

Above the upper bound of the conformal window, the electric theories lose 
asymptotic freedom and become infrared free. Below the conformal window, 
$i.e.$ when $N_f < {3\over 2} N_c \mp {20 \over 3}$, the magnetic theories 
become infrared free and the electric theories are expected to confine. 
This is known as the free magnetic phase. However, we know from planar 
equivalence at infinite $N$ that if we keep reducing the number of flavors, 
at some point the space of vacua of the theory will become empty. 
For $SU(N_c)$ $\NN=1$ SQCD this critical point occurs at $N_f=N_c$, 
precisely when we lose the magnetic description. It is natural to conjecture 
that a similar effect takes place for OQCD at finite $N$. The magnetic 
description is lost at the critical values $N_f=N_c-4$ in OQCD-AS and 
$N_f=N_c+4$ in OQCD-S. Below these values we expect the space of 
vacua of OQCD to be empty. The overall picture is summarized in figure 
\ref{fig:OQCDphases}. For OQCD-AS this picture makes sense only when 
$\frac{3}{2}N_c-\frac{20}{3}>N_c-4$, whereas for OQCD-S it makes sense 
only when $3N_c-\frac{4}{3}>\frac{3}{2}N_c+\frac{20}{3}$. Both inequalities 
require $N_c>\frac{16}{3}$, or equivalently $N_c>5$.

One of the non-trivial effects of the absence of supersymmetry at finite $N$
is the lift of the classical moduli space. To see this, let us consider 
the Coleman-Weinberg potential for the vev of the ``squark'' field 
$\langle \tilde\Phi ^i _\alpha \rangle = \langle \Phi ^i _\alpha \rangle 
\equiv \delta ^i _\alpha v^i$ of the electric OQCD-S theory. Keeping up to 
quadratic terms in the Lagrangian and integrating over the all the
fields, we obtain the effective potential (see ref.\cite{Zarembo:1999hn})
\beq
E(v^1,v^2, \cdots, v^{N_c} ) = N_c \int d^4 p\, \log (p^2 + \sum_{i=1}^{N_c} v^i v^i)- 
(N_c+2) \int d^4 p\, \log (p^2 + \sum_{i=1}^{N_c} v^i v^i) \, ,
\eeq
namely,
\beq
E(v^2) =   -2  \int d^4 p\, \log (p^2 + v^2) \sim -\Lambda ^4 - 
\Lambda ^2 v^2 + (\log \Lambda ^2 / \nu^2 )  v^4 + ... \, , \label{CW}
\eeq
where $v^2 \equiv \sum _i v^i v^i$, $\Lambda$ is a UV cut-off and 
$\nu$ is an IR cut-off. The first term in the above equation~\eqref{CW} 
is a cosmological constant and thus has no implications on the gauge 
theory dynamics. The second term is a mass term for the scalars. It 
may be removed by a fine-tuned renormalization. The one-loop 
generated potential demonstrates that it is impossible to have a quantum 
moduli space at finite $N$. The minimum of the Coleman-Weinberg 
potential is at the origin $v^i=0$, where the electric theory exhibits the 
full $SU(N_f)$ and $U(N_c)$ symmetries. It is worth noting that when the 
``gaugino'' is in the antisymmetric representation (OQCD-AS), $v^i=0$ is 
a maximum of the potential. 

These observations do not invalidate our statement about Seiberg duality. 
It is still possible that there is a duality between the electric and magnetic 
theories in their true (unique) vacuum, rather than a duality in a large moduli 
space.

%%%%%%%%%%%%%%%%%%%%%%%%%%%%%%%%%%%%%%%%%%%
\section{Outlook}

In this paper we considered a non-supersymmetric Seiberg duality between 
electric and magnetic ``orientifold field theories''. These theories are not
generic non-supersymmetric gauge theories. In the large $N$ limit they
exhibit planar equivalence with supersymmetric QCD. This non-trivial statement
gives extra control and allows us to argue convincingly for Seiberg duality
in this limit. Our discussion suggests that the duality may work also at finite $N$. 
An obvious question is whether we can generalize our results in other 
non-supersymmetric cases. Namely, can we find other examples where 
we can argue for non-supersymmetric Seiberg dualities in a similar fashion?
 
Another potential example relies on Seiberg duality between the $SO(N_c)$ 
electric SQCD and the $SO(N_f-N_c+4)$ magnetic theory. To obtain a 
non-supersymmetric descendant, one replaces the adjoint fermion in the 
electric theory by a symmetric fermion. Similarly, one replaces the adjoint 
fermion of the magnetic theory by a symmetric fermion and the symmetric 
fermion in the meson multiplet by an antisymmetric fermion. The result is 
a non-supersymmetric ``electric'' theory and a non-supersymmetric 
``magnetic'' theory. We would like to propose that these theories form a 
pair of Seiberg duals. The evidence for the duality is identical to the one 
in the prime example of this paper. The global anomalies match and 
moreover, we may embed the theory in a non-critical string theory 
version of the Sugimoto model \cite{Sugimoto:1999tx}. In addition, 
the electric and magnetic theories become supersymmetric at large $N$. 
It is interesting to explore this proposal and perhaps others in a future work.

%%%%%%%%%%%%%%%%%%%%%%%%%%%%%%%%%%%%%%%
\medskip
\section*{Acknowledgements}
\noindent
We would like to thank K.\ Hosomichi, E.\ Kiritsis, D.\ Kutasov, S.\ Murthy, A.\ Naqvi, 
C.\ N\'u\~nez, A.~Pakman and J.\ Troost for discussions. A.A.\ is supported by the PPARC 
advanced fellowship award. VN acknowledges partial financial support by 
the INTAS grant, 03-51-6346, CNRS PICS \#~2530, 3059 and 3747 and by 
the EU under the contracts MEXT-CT-2003-509661, MRTN-CT-2004-503369
and MRTN-CT-2004-005104.

%%%%%%%%%%%%%%%%%%%%%%%%%%%%%%%%%%%%%%%
\appendix

%%%%%%%%%%%%%%%%%%%%%%%%%%%%%%%%%
\section{${\mathcal N}$=2 characters and useful identities}
\label{appreps}

In the main text we determine the spectrum of open strings
in various sectors by computing a corresponding set of 
annulus or M\"obius strip amplitudes. The final expressions
of these amplitudes are formulated in terms of the characters 
of the supersymmetric $SL(2)/U(1)$ theory. Such expressions 
appear in appendix \ref{apploop}. In this appendix, we summarize 
the relevant notation and some useful facts about the $SL(2)/U(1)$ 
representations and their characters.

In view of the applications in this paper, we will concentrate on 
the characters of the $SL(2)/U(1)$ super-coset at level $1$. These are 
characters of an $\mathcal{N} =2$ superconformal algebra 
with central charge $c= 9$. They can be categorized in different 
classes corresponding to irreducible representations of the $SL(2)$ 
algebra in the parent WZW theory. In all cases the quadratic Casimir 
of the representations is $c_2=-j(j-1)$. Here we summarize the basic 
representations.

One class of representations are the \emph{continuous representations} 
with $j = 1/2 + ip$, $p \in \mathbb{R}^+$. The corresponding characters 
are denoted by $\chc (p,m) \left[ {a \atop b} \right]$, where the $N=2$ 
superconformal $U(1)_R$ charge of the primary is $Q=2m$, 
$m \in \mathbb{R}$.\footnote{The spectrum of R-charges is not necessarily 
continuous and depends on the model considered. For instance in the cigar 
\textsc{cft} one has $m=(n+w)/2$ with $n,w \in \Z$.} The explicit form of the 
characters is
\begin{equation}
\label{contchar}
\chc (p,m;\tau,\nu)\left[ {a \atop b} \right] =
q^{\frac{p^2+m^2}{k}}e^{4i\pi\nu \frac{m}{k}} \frac{\vartheta\left[ {a \atop b} \right] 
(\tau, \nu)}{\eta^3 (\tau)}\, ,
\end{equation}
where $q=e^{2\pi i\tau}$. This character is similar to the character 
of a free theory comprising of two real bosons and a complex fermion. 
$\vartheta \left[ {a \atop b} \right](\tau,\nu)$ and $\eta(\tau)$ are, 
respectively, the standard theta and eta functions.
 
Another important class of representations comprises of
\emph{discrete representations} with $j$ real. However, none 
of them is normalizable for $k=1$. 

While the closed string spectrum contains only continuous representations, 
the spectrum of open strings attached to localized D-branes is built upon the 
{\it identity representation} with $j=0$. We denote the character of the identity 
representation by $\chid (r)\left[ {a \atop b} \right]$. It has the form
\begin{equation}
\chid (r;\tau,\nu)\left[ {a \atop b} \right] =  \frac{(1-q)\
  q^{\frac{-1/4+(r+a/2)^2}{k}}
e^{2i\pi\nu \frac{2r+a}{k}}}{\left( 1+(-)^b \,
e^{2i\pi \nu} q^{1/2+r+a/2} \right)\left( 1+(-)^b \, e^{-2i\pi \nu}
q^{1/2-r-a/2}\right)} \frac{\vartheta\left[ {a \atop b} \right] (\tau, \nu)}{\eta^3 (\tau)}.
\label{idchar}
\end{equation}
The  primary states in the NS sector, for example, are, besides 
the identity, states of conformal dimension  
\begin{equation}
\Delta =  r^2 -|r| - \frac{1}{2}  \quad , \qquad r \neq 0 ~.
\end{equation}

\subsection*{Extended characters}

When the level $k$ of $SL(2)/U(1)$ is rational it is often convenient to define 
\emph{extended characters} \cite{Eguchi:2003ik}, which serve as
a useful repackaging of ordinary characters. For $k=1$ the extended 
characters are defined by partially summing over integer units of spectral 
flow \cite{Henningson:1991jc,Maldacena:2000hw}. More explicitly, extended 
characters (denoted by capital letters) are defined as
\begin{equation}
{\rm Ch}_\star (\star,\star)\left[ {a \atop b} \right] (\tau;\nu) = \sum_{\ell \in \Z}
{\rm ch}_{\star} \left( \star,\star \right) \left[ {a \atop b} \right] (\tau; \nu +  \ell \tau)
~.
\end{equation}
The stars stand for the appropriate, in each case, representation or quantum
number. For example, the extended characters of the continuous representations 
are for $k$ integer
\begin{equation}
\label{extendcontinuous}
\Chc (P,m)\left[ {a \atop b} \right] (\tau;\nu) = 
\frac{q^{\frac{P^2}{k}}}{\eta^3 (\tau )} \ \Theta_{2m,k} (\tau; 2\nu )
\ \vartheta\left[ {a \atop b} \right] (\tau ; \nu)
~
\end{equation}
with $2 m \in \Z_{2k}$. $\Theta_{n,k}$ is a classical theta function.

\subsection*{Hatted characters}

As usual, M\"obius strip amplitudes are conveniently expressed in
terms of hatted characters. These are defined as
\beq
\label{hatted}
\widehat {\rm ch}_\star(\star;\tau) \left[ {a \atop b} \right]=
e^{-i\pi(\Delta-\nicefrac{c}{24})}{\rm ch}_\star(\star;\tau+\nicefrac{1}{2})
\left[ {a \atop b} \right]
~,
\eeq
where $\Delta$ is the scaling dimension of the primary state of the
representation and $c$ the central charge of the CFT. We refer the 
reader to \cite{Israel:2007si} for a more detailed discussion of the 
properties of these characters in the $SL(2)/U(1)$ super-coset.

\subsection*{Modular transformation properties}

In deriving the one-loop amplitudes in the direct or transverse channel
the following modular transformation properties are useful\footnote{From 
now on we set $\nu=0$.}
\beq
\label{appAaa}
\frac{\vartheta \left[ {a \atop b} \right]\left(-1/\tau\right)}{\eta(-1/\tau)}
=e^{-\frac{i\pi ab}{2}}\frac{\vartheta \left[ {-b \atop a} \right]}{\eta(\tau)}
~, ~ ~ 
\eta(-1/\tau)=(-i\tau)^{1/2} \eta(\tau)
~,
\eeq
\beq
\label{appAab}
\frac{\vartheta \left[ {a \atop b} \right]\left(-\frac{1}{4it}+\frac{1}{2}\right)}
{\eta^3\left(-\frac{1}{4it}+\frac{1}{2}\right)}=
\frac{1}{2t}e^{i\pi(\frac{1}{4}-\frac{b}{2}+\frac{3a}{4})} 
\frac{\vartheta \left[ {a \atop a-b+1} \right](it+\frac{1}{2})}
{\eta^3(it+\frac{1}{2})}
~,
\eeq
\begin{multline}
\label{appAac}
\chc \left(P,m;-\frac{1}{\tau}\right)\left[ {a \atop b} \right]= \\ = 
4 e^{-i\pi ab/2} \int_0^\infty \di P'\int_{-\infty}^\infty \di m'~ e^{-4\pi imm'}
\cos(4\pi PP') \chc (P',m';\tau)\left[ {-b \atop a} \right],
\end{multline}
\begin{multline}
\label{appAad}
\widehat{\chc}(P,m,-\frac{1}{4\tau})\left[ {a \atop b} \right]=
\\
=2 e^{\frac{i\pi}{4}(1-a-2b)} \int_0^\infty \di P' \int_{-\infty}^\infty \di m'~
e^{-2\pi i m m'} \cos(2\pi PP') \widehat \chc(P',m';\tau)\left[ {a \atop a-b+1} \right]
~.
\end{multline}

\subsection*{Relations between identity and continuous characters}

The definition of the extended identity characters is
\beq
\label{appextdef}
\Chid(\tau)\left[ {a \atop b} \right]=\sum_{n\in \Z}
\chid\left(n+\frac{a}{2};\tau\right)\left[ {a \atop b} \right]
~.
\eeq
Some useful identities between continuous and identity characters are 
\beq
\label{appAba}
\Chid(\tau)\left[ {a \atop b} \right]=
\Chc \left(\frac{i}{2},\frac{a}{2};\tau\right)\left[ {a \atop b} \right]-
(-)^b \Chc \left( 0,\frac{1+a}{2};\tau\right)\left[ {a \atop b} \right]
~,
\eeq
\begin{multline}
\label{appAbb}
\sum_{n\in \Z}(-)^n \chid(n;\tau) \left[ {a\atop b} \right] =
- \sum_{n \in \Z} (-)^n \Big\{
\chc \left(\frac{i}{2},n+\frac{a}{2};\tau\right)  \left[ {a\atop b} \right]
\\+(-)^b \chc \left(0,n+\frac{1+a}{2};\tau\right)  \left[ {a\atop b} \right]
\Big\}
~,
\end{multline}
\beq
\label{appAbc}
\sum_{n \in \Z}(-)^n\widehat{\chid}(n;\tau) \left[ {1\atop b} \right]=
\sum_{n \in 2\Z+1}\widehat\chc\left(\frac{i}{2};\frac{n}{2};\tau\right) \left[ {1\atop b} \right] 
+\sum_{n \in 2\Z} (-)^be^{\pi i \nicefrac{n}{2}} \widehat\chc\left(0,\frac{n}{2};\tau\right)
 \left[ {1\atop b} \right]
 ~,
\eeq
\beq
\label{appAbd}
\sum_{n \in \Z}\widehat{\chid}(n;\tau) \left[ {1\atop b} \right]=-
\sum_{n \in 2\Z+1} (-)^\frac{n-1}{2}
\widehat\chc\left(\frac{i}{2};\frac{n}{2};\tau\right) \left[ {1\atop b} \right] 
+\sum_{n \in 2\Z} (-)^b \widehat\chc\left(0,\frac{n}{2};\tau\right)
 \left[ {1\atop b} \right]
 ~.
\eeq

\subsection*{Leading terms in character expansions}

The identity characters have the following leading terms in an expansion 
in powers of $q=e^{2\pi i \tau}$ (here $\tau=it$)

\beq
\label{appAbea}
\sum_{n\in \Z} \chid (n;it)\left[ {0 \atop 0} \right]
\frac{\vartheta \left[ {0 \atop 0} \right](it)}{\eta^3(it)} = 
q^{-\frac{1}{2}}+2+\OO(q)
~,
\eeq
\beq
\label{appAbeb}
\sum_{n\in \Z} \chid (n;it)\left[ {0 \atop 1} \right]
\frac{\vartheta \left[ {0 \atop 1} \right](it)}{\eta^3(it)} = 
q^{-\frac{1}{2}}-2+\OO(q)
~,
\eeq
\beq
\label{appAbec}
\sum_{n\in \Z} \chid (n;it)\left[ {1 \atop 0} \right]
\frac{\vartheta \left[ {1 \atop 0} \right](it)}{\eta^3(it)} = 
4+\OO(q)
~,
\eeq
\beq
\label{appAbed}
\sum_{n\in \Z} (-)^n\widehat\chid (n;it)\left[ {1 \atop 0} \right]
\frac{\vartheta \left[ {1 \atop 0} \right](it)}{\eta^3(it)} = 
4+\OO(q)
~.
\eeq

%%%%%%%%%%%%%%%%%%%%%%%%%%%%%%%%%%%%%%%%%%%%%
\section{One-loop string amplitudes}
\label{apploop}

In this appendix we provide the string theory one-loop amplitudes that are relevant for 
the discussion in the main text.

%%%%%%%%%%%%%%%%%%%%%%%%%%%%
\subsection{The closed string sector}

The starting point of our construction in section \ref{sec:embedding} 
is type 0B string theory on
\beq
\label{closedaa}
\R^{3,1} \times \frac{SL(2)_1}{U(1)}
~.
\eeq
The closed string spectrum of this theory is summarized by the torus
partition function
\bea
\label{closedab}
\TT_{\rm 0B}=&&\frac{V}{2}\int \frac{\di \tau \di \bar \tau}{4\tau_2} \sum_{a,b \in \Z_2}
\frac{\vartheta\left[ {a \atop b} \right](\tau) \vartheta\left[ {a \atop b} \right](\bar \tau)}
{(8\pi^2 \tau_2)^2 \eta^3(\tau)\eta^3(\bar \tau)} \times
\\
&&\sum_{n,w \in \Z_2} \int_0^\infty \di p~ \chc \left(p,\frac{n+w}{2};\tau\right)
\left[ {a \atop b} \right] ~ \chc\left( p,\frac{n-w}{2};\bar \tau\right)\left[ {a \atop b} \right]
~.\nonumber
\eea
In this expression $V$ is a factor that diverges proportionally to
the overall volume of the space-time -- this includes the volume of the
3+1 dimensional flat space-time and the volume of the cigar. The 
continuous representation characters were defined in the previous 
appendix. As far as the fermions are concerned, \eqref{closedab} 
is a typical, 0B diagonal, modular invariant sum.

The Klein bottle amplitude for the orientifold defined in section \ref{closedsector}, 
can be obtained in the transverse channel from the crosscap wave-functions
(\ref{waveorienta}, \ref{waveorientb}). It receives two contributions:
one from the extended part of the crosscap state $|\CC;{\rm ext}\rangle$ and 
another from the localized part $|\CC;{\rm loc}\rangle$. In the direct channel 
(from which we can read off the orientifold action on closed string states), one finds 
by channel duality: 
\begin{subequations}
\bea
\label{closedaka}
&&\KK_{\rm ext}=\frac{1}{2}\int_0^\infty \frac{\di t}{4t^2} \,  \frac{1}{(8\pi^2)^2}
~\langle \mathcal{C};{\rm ext} |e^{-\frac{\pi}{t}H_c} |\mathcal{C};{\rm ext}\rangle=
-\frac{1}{2} \int_0^\infty \frac{\di t}{2t}  \frac{1}{(8\pi^2 t)^2}
\\
~&&\sum_{a\in \Z_2} \sum_{n\in \Z} \int_0^\infty \di p \int_0^\infty \di p'~ 
\frac{\cos(4\pi pp')}{\sinh^2(\pi p)} (-)^{a+n+1} \chc \left(p',\frac{n}{2};2it\right)
\left[ {a \atop 1} \right] \frac{\vartheta \left[ {a \atop 1} \right] (2it)}{\eta^3(2it)}
,\nonumber
\eea
\bea
\label{closedakb}
&&\KK_{\rm loc}=\frac{1}{2}\int_0^\infty \frac{\di t}{4t^2}  \frac{1}{(8\pi^2)^2}
~\langle \mathcal{C};{\rm loc} |e^{-\frac{\pi}{t}H_c} |\mathcal{C};{\rm loc}\rangle=
-\frac{1}{2} \int_0^\infty \frac{\di t}{4t^3}  \frac{1}{(8\pi^2)^2}
\\
&&
\sum_{a\in \Z_2} \sum_{n\in \Z} \int_0^\infty \di p \int_0^\infty \di p'~ 
\cos(4\pi pp') (-)^{a+1} \chc \left(p',\frac{n}{2};2it\right)
\left[ {a \atop 1} \right] \frac{\vartheta \left[ {a \atop 1} \right] (2it)}{\eta^3(2it)}
.\nonumber
\eea
\end{subequations}
The extended contribution to the Klein bottle amplitude is, as expected, divergent
and reproduces the expression anticipated from the known asymptotic form of
the parity \cite{Israel:2007si}. The localized contribution to the Klein bottle amplitude 
exhibits a delta function density of states localized at $p'=0$.

%%%%%%%%%%%%%%%%%%%%%%%%%%%%
\subsection{The open string sectors}

In this paper we consider D-brane setups with D3-branes (the color branes)
and D5-branes (the flavor branes). There are three types of open strings:
color-color strings, flavor-flavor strings and color-flavor strings. 

\subsection*{Color-Color strings}

The annulus amplitude for $N_c$ color D3-branes, characterized by the 
wave-functions~(\ref{wavecoloura}, \ref{wavecolourb}), reads in the open 
string channel
\begin{equation}
\AA_{{\rm D}3-{\rm D}3}= V_{4} N_c^2 
 \int_0^\infty \frac{\di t}{4t}  \frac{1}{(16\pi^2 t)^2} \, 
\sum_{a,b=0}^1  \Chid(it) \left[ {a \atop b} \right] \frac{ \vartheta \left[ {a \atop b} \right](it)}
{\eta^3(it)}
\label{annuluscol}
\end{equation}
where $V_4$ is the volume of $\mathbb{R}^{3,1}$. 
This expression involves only the characters of the identity representation, 
defined in appendix~\ref{appreps}. 

In the presence of the O$'$5 orientifold, the M\"obius strip amplitude is
\begin{multline}
\MM_{D3;{\rm O}'5}=
\mp V_4 N_c 
\int_0^\infty \frac{\di t}{4t} \frac{1}{(16\pi^2 t)^2}
 \sum_{n\in \Z}(-)^n \, \sum_{b=0}^1
\widehat \chid (n;it) \left[ {1 \atop b} \right] 
\frac{\vartheta \left[ {1 \atop b} \right](it+\tfrac{1}{2})}{\eta^3(it+\tfrac{1}{2})}
\, .
\label{mobiuscol}
\end{multline}
The overall sign is fixed by the orientifold charge. For example, the $-$ sign 
leads to a symmetric projection of the gauginos and arises from a positive
orientifold charge. Note that the annulus amplitude~(\ref{annuluscol}) is 
{\it twice} what it would be in type IIB, since the color brane is not a fundamental 
boundary state in type 0B. However, in type 0$'$B this brane is fundamental. 

With the use of these amplitudes and the character expansions 
\eqref{appAbea}-\eqref{appAbed} one can deduce easily the low energy 
degrees of freedom of {\it 3-3} open strings.

\subsection*{Flavor-Flavor strings}

Next we consider the non-compact flavor D5-branes, whose 
wave-functions are given by eqns.~(\ref{waveflavoura}), (\ref{waveflavourb}). 
The annulus amplitude for two flavor branes with parameters 
$(s_1,m_1)$ and $(s_2,m_2)$ reads in the open string channel
\begin{multline}
\AA_{{\rm D}5(s_1,m_1)-{\rm D}5(s_2,m_2)}=  V_4 \int_0^\infty
\frac{\di t}{4t}\frac{1}{(16\pi^2 t)^2} \sum_{a,b=0}^{1} (-)^a 
\frac{ \vartheta \left[ {a \atop b} \right]} {\eta^3}
\int_0^\infty \di P 
\\ \times 
\left\{ \rho_1(P;s_1|s_2) 
 \Chc(P,m_1-m_2+\tfrac{1-a}{2};it) \left[ {a \atop b} \right]\right. \\ \left. 
+ (-)^b \rho_2(P;s_1|s_2) 
\Chc(P,m_1-m_2+\tfrac{a}{2};it)\left[ {a \atop b} \right] 
 \right\}
 ~.
\label{annulusflav}
\end{multline}
In our previous notation, $m_i = \theta_i /2\pi$. 
The spectral densities $\rho_1, \rho_2$ are given by the expressions
\begin{subequations}
\begin{align}
\label{branesala}
\rho_1(P;s_1|s_2)&=4 \int_0^\infty \di P'~
\frac{\cos(4\pi s_1 P')\cos(4\pi s_2 P')\cos(4\pi PP')}
{\sinh^2(2\pi P)}
\, , \\
\label{branesalb}
\rho_2(P;s_1|s_2)&=4 \int_0^\infty \di P'~
\frac{\cos(4\pi s_1 P')\cos(4\pi s_2 P')\cos(4\pi PP')\cosh(2\pi P')}
{\sinh^2(2\pi P)}
\, .
\end{align}
\end{subequations}
Both densities have an infrared divergence at $P'\to 0$, which is related to the 
infinite volume of the cigar. This infinity can be regularized by computing 
{\it relative} annulus amplitudes, $i.e.$ by substracting the annulus amplitude of
a reference brane \cite{Ponsot:2001gt,Fotopoulos:2005cn}.

In section \ref{sec:embedding} we consider D5-branes with $\theta=0 \mod \pi$.
The setup of subsection \ref{electricsetup} involves $N_f$ D5 boundary
states $\overline{|D5(s,\nicefrac{1}{2}\rangle}$ for which the M\"obius strip 
amplitude reads in the open string channel
\begin{multline}
\label{moduliaa}
\MM_{\overline{{\rm D}5(s,\nicefrac{1}{2})};{\rm O}'5}=
\mp N_f \int_0^\infty \frac{\di t}{4t} \frac{1}{(16\pi^2 t)^2}
 \int_0^\infty \di P   \\ \sum_{N \in \mathbb{Z}} \left[ 
(-)^N \rho_3 (P;s)\, \widehat \chc(P,N;it)\left[ {1 \atop 0} \right]
\frac{\vartheta \left[ {1 \atop 0} \right](it+\nicefrac{1}{2})}{\eta^3(it+\nicefrac{1}{2})}\right. \\ - 
\left.  \rho_4 (P;s) \, 
\widehat \chc(P,N+\tfrac{1}{2};it)\left[ {1 \atop 0} \right]
\frac{\vartheta \left[ {1 \atop 0} \right](it+\nicefrac{1}{2})}{\eta^3(it+\nicefrac{1}{2})} \right] 
\end{multline}
with spectral densities
\begin{subequations}
\begin{align}
\rho_3 (P;s) &= 2\int_0^\infty\di P'\, \frac{\cos(8\pi sP')\cos(4\pi PP')}{\sinh^2(2\pi P')} 
= \rho_1 (P;s|s) - 2\int_0^\infty\di P\,' \frac{\cos(4\pi PP')}{\sinh^2(2\pi P')} \, ,\\
\rho_4 (P;s) &= 2\int_0^\infty 
\di P' \, \frac{\cos(8\pi sP)\cos(4\pi PP')}{\cosh(2\pi P)}
\, . 
\end{align}
\end{subequations}
We observe that the density $\rho_3$, coming from the 
extensive part of the crosscap, is divergent, and coincides with the 
density $\rho_1$ in the annulus amplitude for two identical flavor branes, 
see eq.~(\ref{branesala}), up to an $s$-independent term that cancels out 
when we compute a {\it relative amplitude} in order to remove the infrared 
divergence. The density $\rho_4$ is infrared finite, as it comes from the 
localized piece of the orientifold. 

The low-energy spectrum of {\it 5-5} open strings in section 
\ref{sec:embedding} can be determined straightforwardly from the above
amplitudes.

\subsection*{Color-Flavor open strings}

Flavor degrees of freedom in gauge theory arise from {\it 3-5} strings.
The open string spectrum of this sector can be determined from 
the annulus amplitude between a color D3-brane and a flavor D5-brane.  
By definition, there is no M\"obius strip amplitude contributing to this sector. 
The flavor brane has parameters $(s,\theta)$. In order to cover the general case 
we will include here amplitudes involving flavor branes with both positive and 
negative RR-charge, $i.e.$ both $D5$ and $\overline{D5}$ branes.
We call $\varepsilon=\pm 1$ the sign of the flavor RR-charge. 
One  finds in the open string channel:
\begin{equation}
\label{annul35}
\AA_{{\rm D}3-{\rm D}5(s,\theta,\epsilon)}=
\int_0^\infty \frac{\di t}{4t} \frac{1}{(16\pi^2 t)^2}\sum_{a,b=0}^1 (-)^{a+b(1-\varepsilon)}
 \Chc(s,\tfrac{\theta}{2\pi} +\tfrac{a}{2};it) \left[ {a \atop b} \right] 
\frac{ \vartheta \left[ {a \atop b} \right](it)}{\eta^3(it)}
~.
\end{equation}

There are no non-trivial spectral densities in this case and one can read off
the spectrum immediately by using the extended continuous character definition
\eqref{extendcontinuous}.

%%%%%%%%%%%%%%%%%%%%%%%%
\section{RR charges}
\label{app:RRcha}

In this appendix we determine the RR scalar charge 
of the D3-, D5-branes and of the O$'$5-plane. 
The massless RR field $C_0$ has quantum numbers
$P=0,w=0,n=0$. The charge of the above objects 
is proportional to the one-point of $C_0$ on the disc.
This quantity is provided by the wave-functions of the
boundary/crosscap states in the main text.

Specifically, for D3-branes we have (see eq.\  \eqref{wavecolourb}):
\beq
\label{appCaa}
\lim_{P\to 0} \Phi_R(P,0)=\NN_{{\rm D}3} \lim_{P \to 0}
\frac{\Gamma(iP)}{\Gamma(2iP)}=2\NN_{{\rm D}3}
~.
\eeq
For D5-branes of the type $\overline{|D5;0,\frac{1}{2}\rangle}$
we have (see eq.\ \eqref{waveflavourb}):
\beq
\label{appCab}
\lim_{P\to 0} \Phi_R\left(\overline{0,\frac{1}{2}};P,0\right)=-\NN_{{\rm D}5}
\lim_{P\to 0}\frac{\Gamma(-2iP)}{\Gamma(-iP)}=-\frac{\NN_{{\rm D}5}}{2}=-
\NN_{{\rm D}3}
~.
\eeq
Finally, for the O$'$5-plane (see eq.\ \eqref{waveorientb})
\beq
\label{appCac}
\lim_{P\to 0}(P,0;+)=\NN_{{\rm O}'5} \lim_{P\to 0}
\frac{\Gamma(-2iP)}{\Gamma(-iP)}=\frac{\NN_{{\rm O}'5}}{2}=
4\NN_{{\rm D}3}
~.
\eeq
In these expressions $\NN_{{\rm D}3}$, $\NN_{{\rm D}5}=2\NN_{{\rm D}3}$ 
and $\NN_{{\rm O}'5}=8\NN_{{\rm D}3}$ are respectively the normalizations 
of the D3, D5 boundary states and the O$'$5 crosscap state.

%%%%%%%%%%%%%%%%%%%%%%%%%%%%%%%%%
\section{Forces between D-branes and O-planes}
\label{appMob}

Here we consider the forces that arise between the branes 
and/or orientifold planes at one string loop in our setups. By 
construction, the annulus amplitudes vanish, hence they do
not give a net potential to the brane moduli. The M\"obius 
strip amplitudes, however, get contributions from the RR 
sector only and break the Bose-Fermi degeneracy of the open
string spectra. This breaking of supersymmetry leads to attractive 
or repulsive forces between the O$'$5 orientifold plane and the 
D-branes. Since the color D3-brane has no moduli, this contribution 
generates only a constant potential. However, the M\"obius strip 
amplitude for the flavor D5-brane generates a potential 
for the brane modulus $s$, which characterizes the minimum
distance of the brane from the tip of the cigar. 

From the closed string point of view, the force between the orientifold 
and the flavor brane comes from the exchange of  massless RR closed 
string modes. It is captured by the $t\to 0$ (or $\ell\equiv 1/t \to \infty$) 
asymptotics of the M\"obius strip amplitude~(\ref{moduliaa}):
\begin{equation}
\label{moduliab}
\MM_{\overline{{\rm D}5(s,\nicefrac{1}{2})};{\rm O}'5}\sim\pm N_f
  \int^\infty \di \ell  \int_0^\infty \di P\, 
 \frac{\cos(4\pi sP)}{\cosh (\pi P)} e^{-\frac{\pi \ell}{2} P^2} \left[ 1+ \mathcal{O}
\left(e^{-\tfrac{\pi \ell}{2}}\right) \right]_{RR} ~.
\end{equation}
Only the localized part of the orientifold sources massless fields, hence 
this is the only one that contributes in this regime. We are interested 
in the small $s$ behavior of the function 
$F(s)\equiv -\partial_s \MM_{\overline{{\rm D}5(s,\nicefrac{1}{2})};{\rm O}'5}$,
where quarks from the {\it 3-5} sector are nearly massless. $F(s)$ is the 
``force'' felt by the flavor branes in the presence of the orientifold
plane. Since we focus on the $\ell \to \infty$ asymptotics, the $P$-integral in  
\eqref{moduliab} is dominated by the region $P\to 0$. Hence, we get a linear 
expression in the modulus $s$ at first order: $F(s)\propto \pm\, s$. This result 
shows that the force is attractive towards $s=0$ for an orientifold with positive 
charge (OQCD-S setup). Similarly, for an orientifold with negative charge (the 
OQCD-AS setup) the force is repulsive.

It should be noted that this force, which arises as a one-loop effect on the 
flavor branes, has no consequences on the dynamics of the four-dimensional 
OQCD theory that was engineered in the intersection of D3- and D5-branes 
in section \ref{sec:embedding}. In order to isolate the four-dimensional gauge 
theory dynamics from the rest of the open string degrees of freedom we 
need to take a low-energy decoupling limit where the dynamics on the 
flavor D5-branes are frozen. In this limit, only fields from the {\it 3-3} and 
{\it 3-5} sectors can run in loops to generate potentials. The force $F(s)$
appearing above is not one of these effects, because it arises from a loop 
in the {\it 5-5} sector.

%\newpage
\addcontentsline{toc}{section}{References}

%\bibliographystyle{utphys}
%\bibliography{bibOQCD}

\end{document}